\documentclass[fleqn,usenatbib]{mnras}

\usepackage{newtxtext,newtxmath}

\usepackage[T1]{fontenc}

\DeclareRobustCommand{\VAN}[3]{#2}
\let\VANthebibliography\thebibliography
\def\thebibliography{\DeclareRobustCommand{\VAN}[3]{##3}\VANthebibliography}

\usepackage{caption,float}
\usepackage{graphicx}	
\usepackage{amsmath}	

\def\xs{M106}
\def\sou{the eROSITA/Fermi bubbles}
\def\rins{LOFAR}

\title[\xs\ bubbles]{Tracing the Energetic Outflows from Galactic Nuclei: Observational Evidence for a Large-Scale Bipolar Radio and X-ray-emitting Bubble-like Structure in M106}

\author[Yuxuan Zeng et al.]{Yuxuan Zeng$^{1}$, Q. Daniel Wang$^{2}$\thanks{Contact email:wqd@umass.edu}, Filippo Fraternali$^{1}$\\
$^{1}$ Kapteyn Astronomical Institute, University of Groningen, PO Box 800, 9700 AV Groningen, The Netherlands\\ 
$^{2}$ Department of Astronomy, University of Massachusetts, Amherst, MA 01003, USA \\
}

\pubyear{2023}

\begin{document}
\label{firstpage}
\pagerange{\pageref{firstpage}--\pageref{lastpage}}
\maketitle




\begin{abstract}
The role of energetic outflows from galactic nuclei in shaping galaxy formation and evolution is still shrouded in uncertainty. In this study, we shed light on this complex phenomenon by presenting evidence for a large-scale bipolar radio/X-ray-emitting bubble-like structure emanating from the central region of the nearby disk galaxy M106 (NGC\,4258). Our findings, based on Low-Frequency Array survey data and Chandra observations, provide a glimpse into the underlying physical processes driving this enigmatic structure. Similar to the eROSITA/Fermi bubbles in our own Galaxy, the M106 bubbles enclose diffuse hot plasma and are partially bounded by prominent radio/X-ray-emitting edges. We constrain the magnetic field and cosmic-ray properties of the structure. The analysis of the X-ray data gives an estimate of the thermal energy of the bubbles as $\sim 8 \times 10^{56}$ erg. This energy can be supplied by the jets and perhaps by the wind from the accretion flow of the galaxy’s low-luminosity AGN, which most likely has been much more powerful in the recent past, with an average mechanical energy release rate of $\sim 4 \times 10^{42} {\rm~erg~s^{-1}}$ over the last $\sim 8 \times 10^6$~yr -- the estimated age of the structure. 
We also show evidence for diffuse X-ray emission on larger scales, indicating the presence of a hot galactic corona. Our results provide a clear manifestation of galactic nuclear feedback regulating the gas content and energetics of the circumgalactic medium of disk galaxies similar to our own. 
\end{abstract}
\begin{keywords} galaxies: jets, halos, ISM: general, radio continuum: ISM, X-rays: general, ISM\end{keywords}
\section{Introduction}
Galactic feedback, in the form of supernovae and active galactic nuclei (AGN), plays a crucial role in the modern theory of galaxy formation and evolution. It is predicted to have significant effects on galaxies, such as quenching star formation, regulating the growth of supermassive black holes (SMBHs), and driving the circulation of the interstellar and circumgalactic medium of different phases and metallicities \citep{Dekel2006,Martig2009,Fabian2012,Hopkins2012,Tumlinson2017,Li2013,Li2017}. However, much remains uncertain about the effectiveness of feedback or its coupling to the medium \citep[e.g.,][]{Pillepich2021,Schellenberger2023,Truong2023}.

Even the origin of such prominent galactic structures as the well-known Fermi and eROSITA bubbles observed in our Galaxy \citep{bland-hawthorn2003,Su2010,bland-hawthorn2019,Predehl2020} remains a subject of debate \citep[e.g.,][]{Pillepich2021,Yang2022,Sarkar2023}. The Fermi bubbles, first detected in $\gamma$-ray in 2010, extend over 10~kpc above and below the Galactic plane and appear to be associated with a bipolar diffuse X-ray feature observed toward the Galactic central field \citep{Wang2002,bland-hawthorn2003,bland-hawthorn2019}. This connection became more apparent a decade later when the eROSITA all-sky survey revealed similar bubbles in X-rays, known as the eROSITA bubbles, surrounding the Fermi bubbles and extending out to 14 kpc from the Galactic plane. Parts of this $\gamma$-ray/X-ray structure have apparent radio counterparts \citep[e.g., Radio Loop I;][]{Berkhuijsen1971,Carretti2013}, although their physical association is still uncertain, largely due to the severe projection confusion of interstellar features along the long sightline through the Galactic disk toward the Galactic center \citep[e.g.,][]{Das2020,Panopoulou2021}. 

One scenario for the origin of these structures is that they result mainly from supernova feedback, in which shock-heated gas expands from the central Galactic region \citep[e.g.,][]{Crocker2011,Carretti2013,Sarkar2019}. 
Alternatively, the bubbles could be produced by jets or other forms of energetic outflows from Sgr A* in the recent past \citep{Guo2012,Zubovas2012,Yang2022}. However, many uncertainties remain in both the modeling of these scenarios and the interpretation of the observations \citep[e.g.,][]{Miller2016,Nogueras-Lara2020,Yang2022,Sarkar2023,Gupta2023}. As a result, the origin of the eROSITA/Fermi bubbles is still unclear \citep[e.g.,][]{Kataoka2018}.
It is thus highly desirable to get clues from studying similar structures in and around nearby disk galaxies.

\begin{figure}
  \includegraphics[width=\columnwidth]{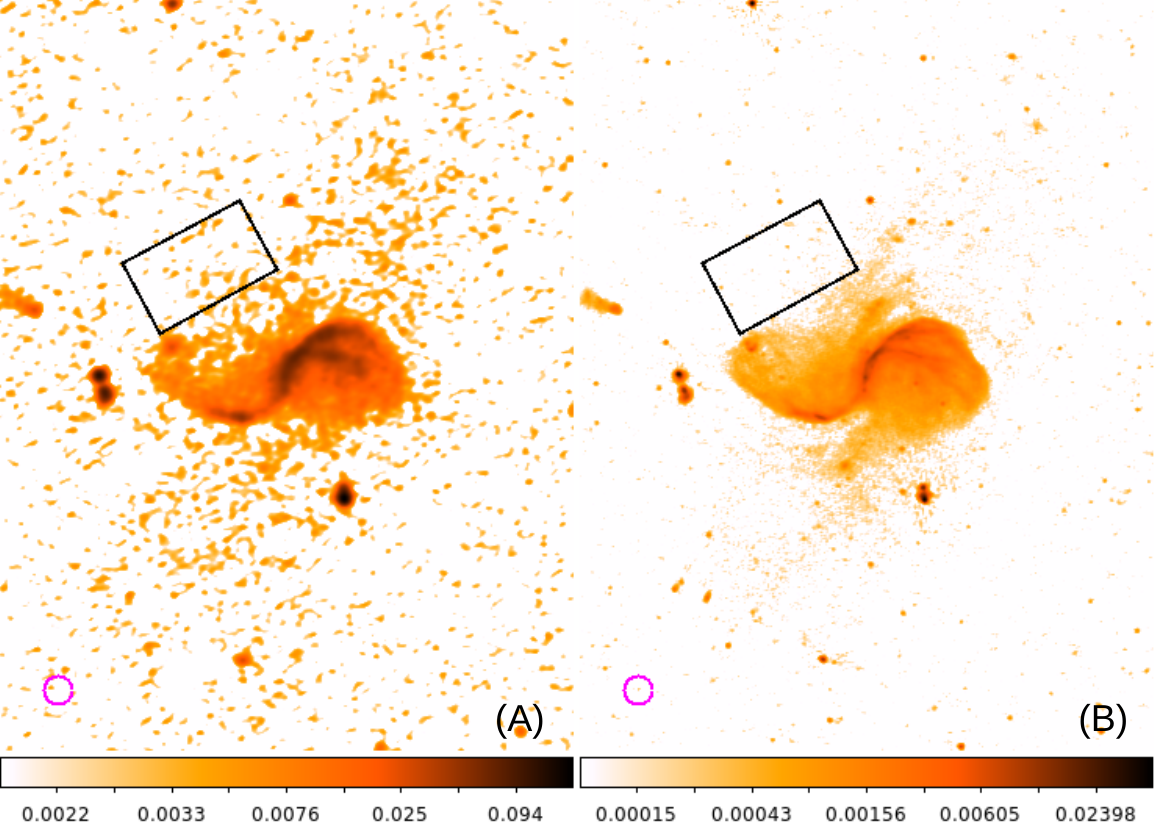}
  \caption{\rins\ intensity images of M106 in units of Jy~beam$^{-1}$ at 54 MHz (left) and 144 MHz (right). The magenta circle in the lower left corner has a 1 kpc radius at the distance of the galaxy, while the outlined rectangular region is used to estimate the background and its RMS. The logarithmically scaled color bars are optimized to show the large-scale diffuse radio emissions. The resolution for 54 MHz is $15^{\prime\prime}$, while for 144 MHz it is $6^{\prime\prime}$.}
  \label{fig:f1}
\end{figure}

\begin{figure*}
    \centering
    \includegraphics[width=1\textwidth]{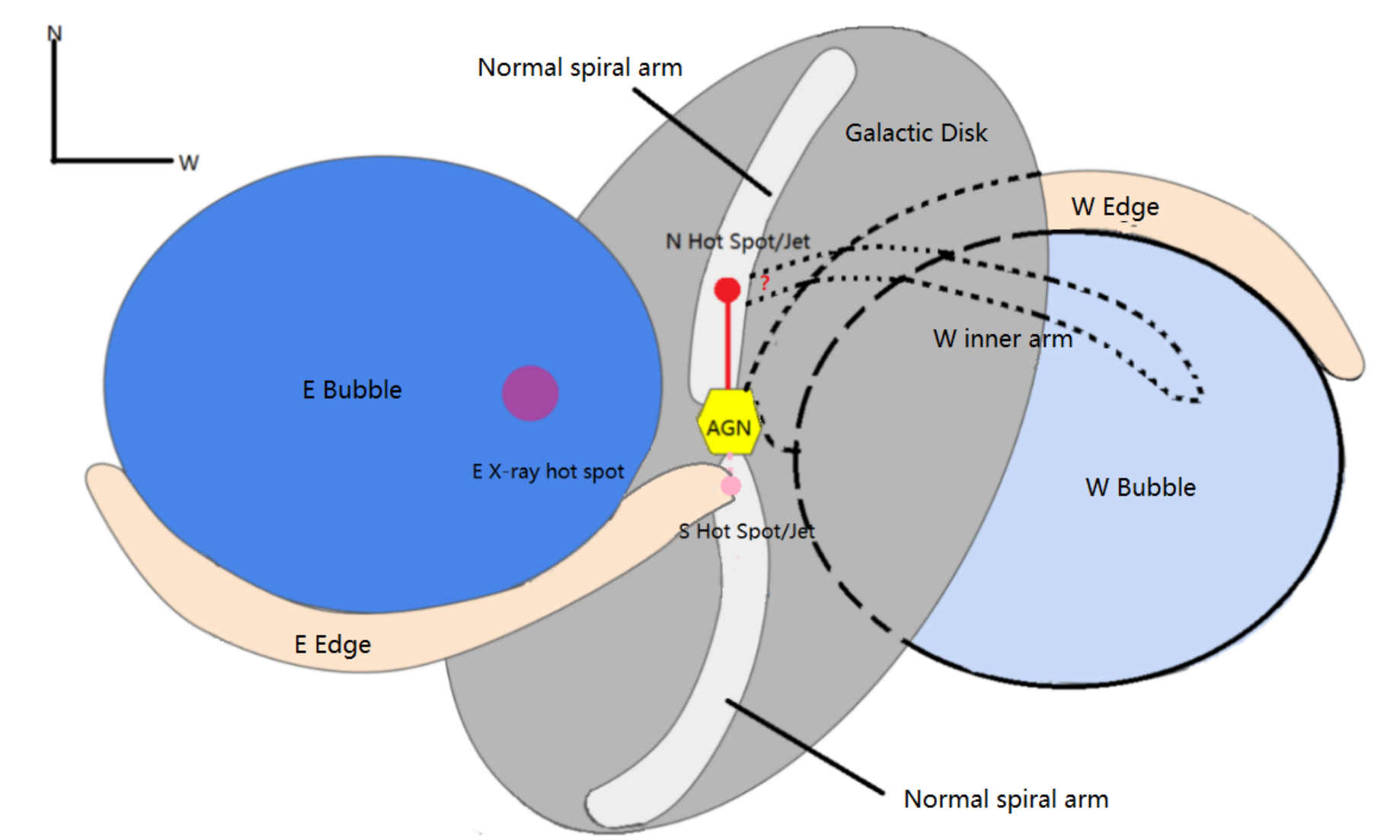}
\caption{Illustration of the main components of \xs\ concerned in the present work: the galactic disk (grey ellipse) as seen in optical; the east and west radio/X-ray emitting bubbles reported here, consisting of both their interiors (represented by the two blue ellipses) and their outer edges, previously known as the two major anomalous arms (marked here as E and W edges). The inclined disk has its near side in front of the western part of this bipolar superbubble structure, apparently originating from the AGN, which is also marked together with two radio-observed jets. We also marked the inner anomalous arm in the Western bubble (W inner arm). This structure could extend to the Northern hot spot and be associated with it although both extension and association are very uncertain (hence the question mark; see later discussion (Section \ref{s:res}) and Fig.~\ref{fig:f8}).}
  \label{fig:f2}
\end{figure*}

We herein present the detection and study of a bipolar superbubble structure, apparently driven by an AGN, in M106, using both Chandra X-ray and Low Frequency Array (\rins) radio data (Fig.~\ref{fig:f1}). Table~\ref{t:t1} lists the salient parameters of this nearby disk galaxy, which is very similar to our own. 
At the distance of \xs, $1^\prime = 2.22$~kpc.
For ease of reference, Fig.~\ref{fig:f2} illustrates the main components of the galaxy that are of interest in this paper. The galaxy is known for the presence of bright "anomalous" arms that are significantly offset from the normal spiral arms \citep{Courtes1960,vanderKruit1972} and have been detected in radio, H${\alpha}$, and X-ray observations \citep[e.g.,][]{Hummel1989,Cecil2000,Yang2007}. Extensive studies have been carried out on these anomalous arms, as well as on the low luminosity AGN and its jets \citep[e.g.,][]{Makishima1994,Lasota1996,VronCetty2006,Masini2022}, the normal spiral arms, and the Galactic disk \citep{Laine2010,Ogle2014}. The anomalous arms are thought to be produced by the jets, which point in directions quite different from the orientation of the superbubble structure. The jets have deposited much of their energy in the ambient medium, probably via fast precession through the galactic disk of the galaxy 
\citep[]{Cecil2000,Yang2007}. 

In this work, we interpret the anomalous arms as the southern and northern bright parts of the outer boundaries of the radio/X-ray east (E) and west (W) bubbles (Figs.~\ref{fig:f2}-\ref{fig:f3}). These brightened edges of the two bubbles are hereafter referred to as the E and W edges. The presence of the bubbles is also apparent in some of the existing VLA data of the galaxy \citep[e.g.,][]{Sofue1980,Cecil1995,Wilson2001}, although they have never been specifically studied, in particular in a multi-wavelength context.
Such unilaterally enhanced radio/X-ray edges are also present in or around the eROSITA/Fermi bubbles, although line-of-sight confusion with other features in the Galactic disk has prevented a firm physical association \citep[e.g.,][]{Das2020,Panopoulou2021}. With the moderate inclination angle of the \xs\ disk (Table~\ref{t:t1}) such confusion is small. So the physical association of the anomalous arms with the bubbles is quite clear. 
Those anomalous arms or features projected inside the bubbles are typically fainter, except for the W inner arm, the brightness of which is comparable to that of the W edge. They appear to result from the bifurcation of the flows driven by choked jets \citep[e.g.,][]{Hummel1989,Krause2004,Sarkar2023}. 
In short, the large-scale bipolar superbubble structure of \xs\ is an excellent case for studying the interplay of galactic nuclear outflows with the interstellar medium (ISM) and the circumgalactic medium (CGM) in a nearby disk galaxy.

The organization of this paper is as follows: We describe the reduction and analysis of the radio and X-ray data in section \ref{s:reduction} and present our results in section \ref{s:res}. In section \ref{s:dis}, we discuss the implications of our results in comparison with similar features observed in other galaxies, in particular, \sou, and with cosmological simulations. Finally, in section \ref{s:sum}, we summarize the main findings of this work.

\begin{table}
	\caption{Parameters of M106}
\begin{tabular}{lr}
\hline
\hline   
Parameter                       & Values\\
\hline   
Galaxy Name &M106, NGC4258\\  
Type                            &SABc\\    
Distance (Mpc)                  & 7.6  \\
M$_B$ (mag) & 20.59\\
$M_* (10^{10} {\rm~M_\odot}$) & 8.2 \\
SFR (${\rm M_\odot~yr^{-1}}$) &1.4\\
Disk incl. ($\deg$)                  & $71^{\circ}$  \\
Disk rotation $({\rm~km~s^{-1}})$& 208\\
${\rm N_{\rm H, G}} (10^{20}~{\rm cm^{-2}}$) & 4.21    \\
\hline 
\end{tabular}	
\label{t:t1}\\
Note: Parameters are obtained from NED/SIMBAD, except for the Type and magnitude from \cite{2011Heald}, stellar mass ($M_*) $ from \cite{Burbidge1963},  foreground Galactic HI column density (${\rm N_{\rm H,G}}$) from \cite{HI4PICollaboration2016}, and star formation rate (SFR) from \cite{Ogle2014}.
\end{table}

\section{Data reduction and analysis}\label{s:reduction}
\subsection{\rins\ data}\label{s:lof}

The present work began with an examination of the recently released \rins\ data. The data cover \xs\ in both the 41-66 MHz band (with an effective center frequency of $\nu_1=54$~MHz and a resolution of 
 $15^{\prime\prime}$ FWHM) and the 120-168 MHz band ($\nu_2=144$~MHz and $6^{\prime\prime}$) from the \rins\ Low Band Antenna Sky Survey (LoLSS) DR1 \citep{deGasperin2023} and the \rins\ Two Meter Sky Survey (LoTSS) DR2 \citep{Shimwell2022}, respectively.
The released survey data (Fig.~\ref{fig:f1}) are of sufficiently high quality to study both the overall morphology and the intensity distributions of the substructures.

We further reduce the data to map the spectral index of the radio emission. First, we convolve the 144 MHz image to the resolution of the 54 MHz image. Second, we remove from each image a local background, which is the median intensity in a "clean" neighborhood of \xs\ (Fig~\ref{fig:f1}): 1.6 mJy\,beam$^{-1}$ at 54~MHz or 0.54 mJy\,beam$^{-1}$ at 144~MHz. This background subtraction does not generate any significant effect here but is applied anyway for consistency since it is also used in the X-ray data analysis to remove the potential large-scale halo contribution of the galaxy. 
Third, the root mean square (RMS) of the intensity in the region, 4.4 or 1.1 mJy\,beam$^{-1}$, is used as the empirical noise estimate for the background subtracted 54 or 144 MHz image. 
Fourth, only the field with a signal-to-noise ratio (S/N) greater than 3 in both images, further excluding regions contaminated by discrete compact radio sources, is retained for the calculation of the spectral index, which is defined as
\begin{equation}\label{eq. alpha}
  \alpha = -\frac{\ln(S_2/S_1)}{\ln(\nu_2/\nu_1)},
\end{equation}
where $S_1$ and $S_2$ are the intensities at the two frequencies.
When calculating the average spectral index of a region, we estimate its error as the RMS divided by the square root of the number of covered beams (FWHM$=15^{\prime\prime}$).
We also use the VLA 8.44 GHz data described by \cite{Krause2004} to calculate the spectral index with LOFAR 144 MHz and present detailed results in Section \ref{s:res}.

\begin{figure*}
  \includegraphics[width=1\textwidth]{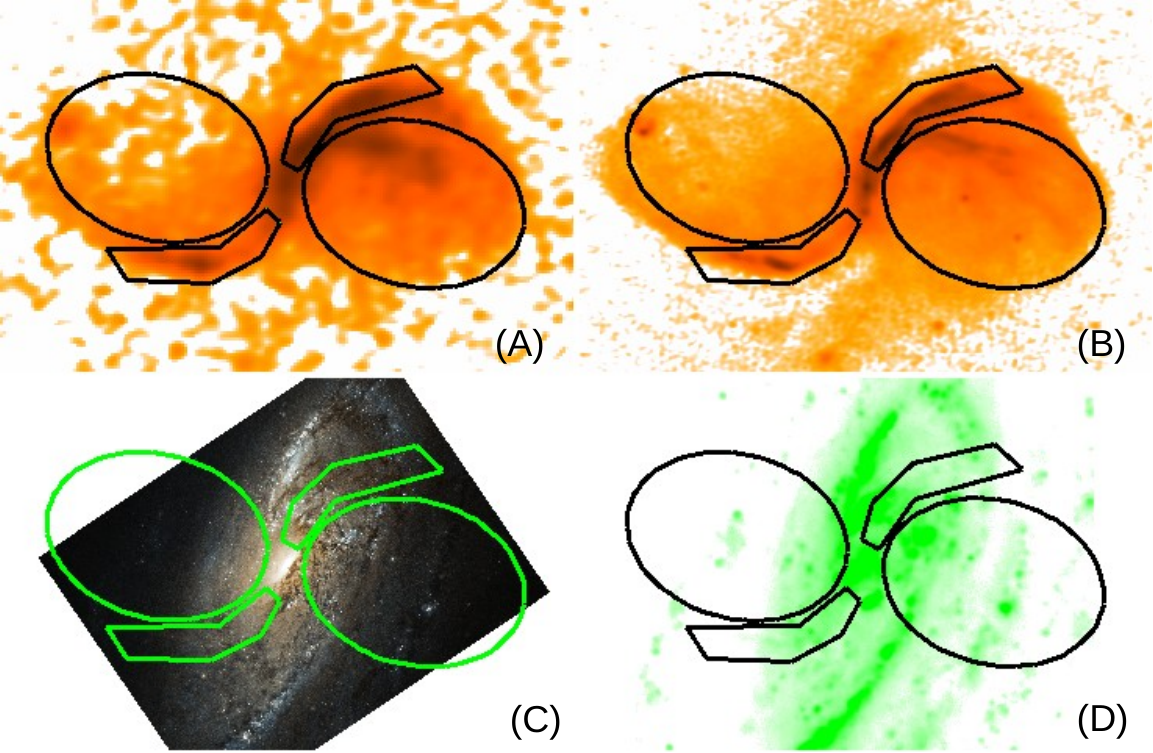}
\caption{Radio detection of the bipolar superbubble structure in \xs: (A) LoLSS DR1 54 MHz and (B) LoTSS DR2 144 MHz images of the galaxy. The interiors of the bubbles are characterized by the two ellipses with a semi-major/minor axis of 4/3 kpc, while parts of their outer edges are outlined for our spectral analysis of enhanced X-ray emission (see also Fig.~\ref{fig:f5}). In addition, the region of the W inner arm in the W bubble is also outlined. For comparison, these outlined regions are also shown in (C) the 3-color image obtained in the three HST filters (red - f814w, green - f555w, and blue - f438w) and (D) the GALEX FUV image of the galaxy. 
}
  \label{fig:f3}
\end{figure*}

\begin{figure*}
  \includegraphics[width=1\textwidth]{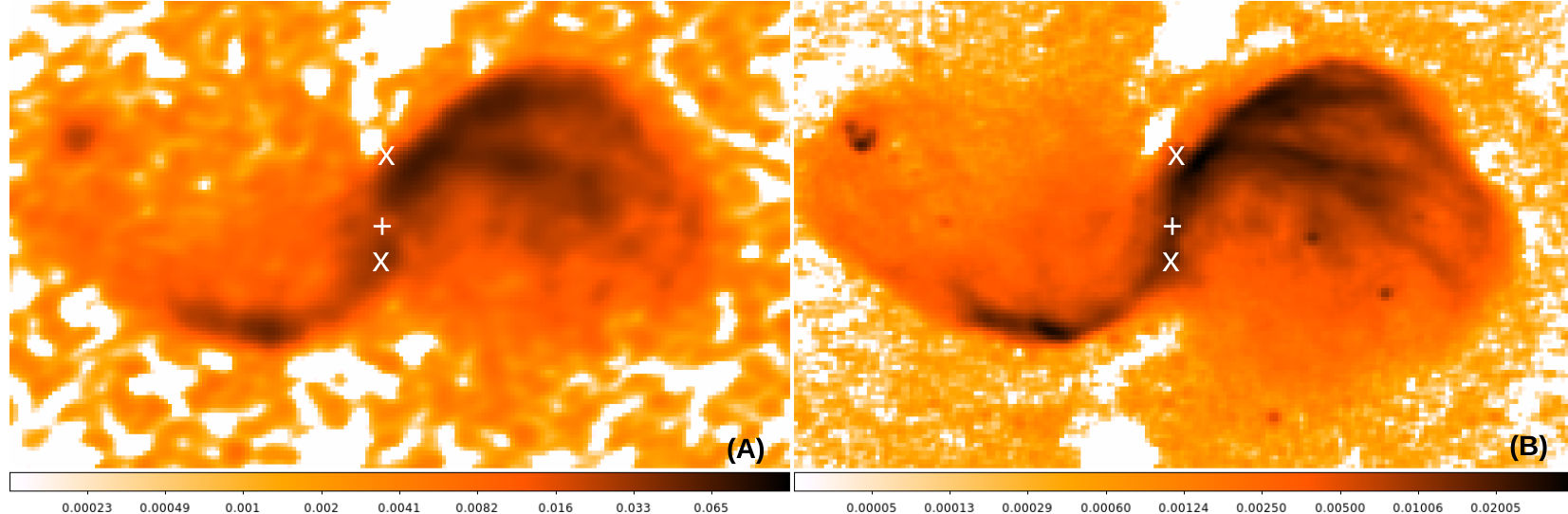}
\caption{Same as Fig.~\ref{fig:f3}A and B, but with the disk contributions approximately subtracted. The positions of the AGN and the N and S radio hot spots are marked by + and Xs, respectively. 
}
  \label{fig:f4}
\end{figure*}

We also try to isolate the radio emission of the galactic disk from that of the bipolar structure. This is especially useful for a more accurate calculation of the spectral index of the radio emission from the \xs\ structure. To do this, we use a WISE 22$\,\mu m$ intensity image of the galaxy to trace the radio contribution from the galactic disk. The image, downloaded from the InfraRed Science Archive (IRSA) \footnote{https://irsa.ipac.caltech.edu/Missions/wise.html}, contains a strong background. We estimate it in the same off-galaxy field as marked in Fig.~\ref{fig:f1} and subtract it from the whole image. The resulting net 22$\,\mu m$ emission from the galaxy should be mostly due to dust-reprocessed UV radiation from massive stars. Their feedback is also expected to be responsible for the acceleration of the cosmic ray particles producing much of the radio emission from the disk. The WISE 22$\,\mu m$ intensity may not exactly follow the synchrotron radiation, because the underlying diffusion or transfer processes of cosmic ray particles and UV radiation may be quite different. However, we find that an approximate subtraction of the disk contribution is sufficient to test its effect on the radio index calculation. In each of the two \rins\ bands, we adjust the ratio of the radio to 22$\,\mu m$ intensity so that the radio image looks uniform over the disk regions after subtracting the disk contribution (Fig~\ref{fig:f4}). The ratios of the two \rins\ bands are then used to estimate the mean spectral index ($1.00\pm 0.20$) of the disk. With the resulting disk-subtracted images, we rebuild the radio spectral index map. We find that the effect on the radio index calculation is small.

\subsection{Chandra data}\label{s:obs}
\subsubsection{Data selection and calibration}\label{sec:dataselect}

\begin{figure}
  \includegraphics[width=\columnwidth]{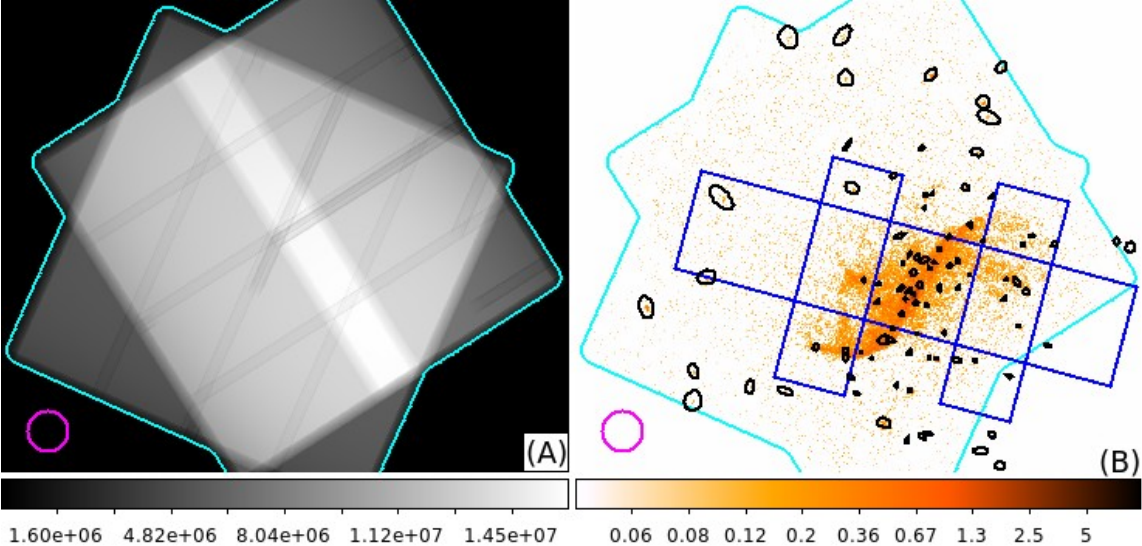}
  \caption{Overview of the Chandra data used in the present study of M106: (A) the effective exposure map (in units of s cm$^{2}$) constructed in the 0.45-1 keV band; (B) the 0. 5-7 keV count map, together with the black ellipses enclosing the 90\% EER regions of individual detected sources and the boxes outlining the regions used to construct the 1-D radio/X-ray intensity distributions shown in Fig~\ref{fig:f12} (the regions parallel and perpendicular to the galaxy's minor axis have dimensions of 5.5 kpc $\times$ 23.7 kpc and 3.6 kpc $\times$ 11.5 kpc). In both panels, the field covered by the data is outlined by the cyan contour at the effective exposure of $1\times10^6$ s cm$^2$, while the magenta circle in the lower left corner has a 1 kpc radius at the distance of the galaxy. }
  \label{fig:f5}
\end{figure}

Our X-ray study of \xs\ uses the same two Chandra observations described in the work by \citet{Yang2007} (see also Table~\ref{t:t2} ). While this early work is focused on the anomalous arms, our study here is interested chiefly in the large-scale diffuse X-ray emission associated with the radio bubbles and its relation to the arms. These observations were made with the Advanced CCD Imaging Spectrometer-Spectroscopy (ACIS-S). We use only the data collected by the S3 CCD chip, which covers M106. Fig.~\ref{fig:f5}A shows the effective exposure map of the combined data. We use the Chandra Interactive Analysis of Observations (CIAO) software (version 4.14 with CALDB 4.9.7) to process the data, following the standard procedure which includes the cleaning of time intervals with strong background flares using the DEFLARE tool, and the merging of the count and exposure images to produce the mosaic maps in the 0.45-1, 1-2, and 2-7 keV bands, as well as the broad (0.45-7 keV) band. 
\begin{table}
	\centering
	\tabcolsep=0.2cm 
\	\caption{Chandra Observation details of M106}
	\label{t:t2}
	\begin{tabular}{lcccr} 
		\hline
		Obs\_ID &Cleaned Exposure& Mode & Dates\\
		&ks&\\
		\hline
		350&14.04&FAINT&2000-04-17 \\
		1618&20.92&VFAINT&2001-05-28\\
		\hline
	\end{tabular}
\end{table}

\begin{figure}
\includegraphics[width=\columnwidth]{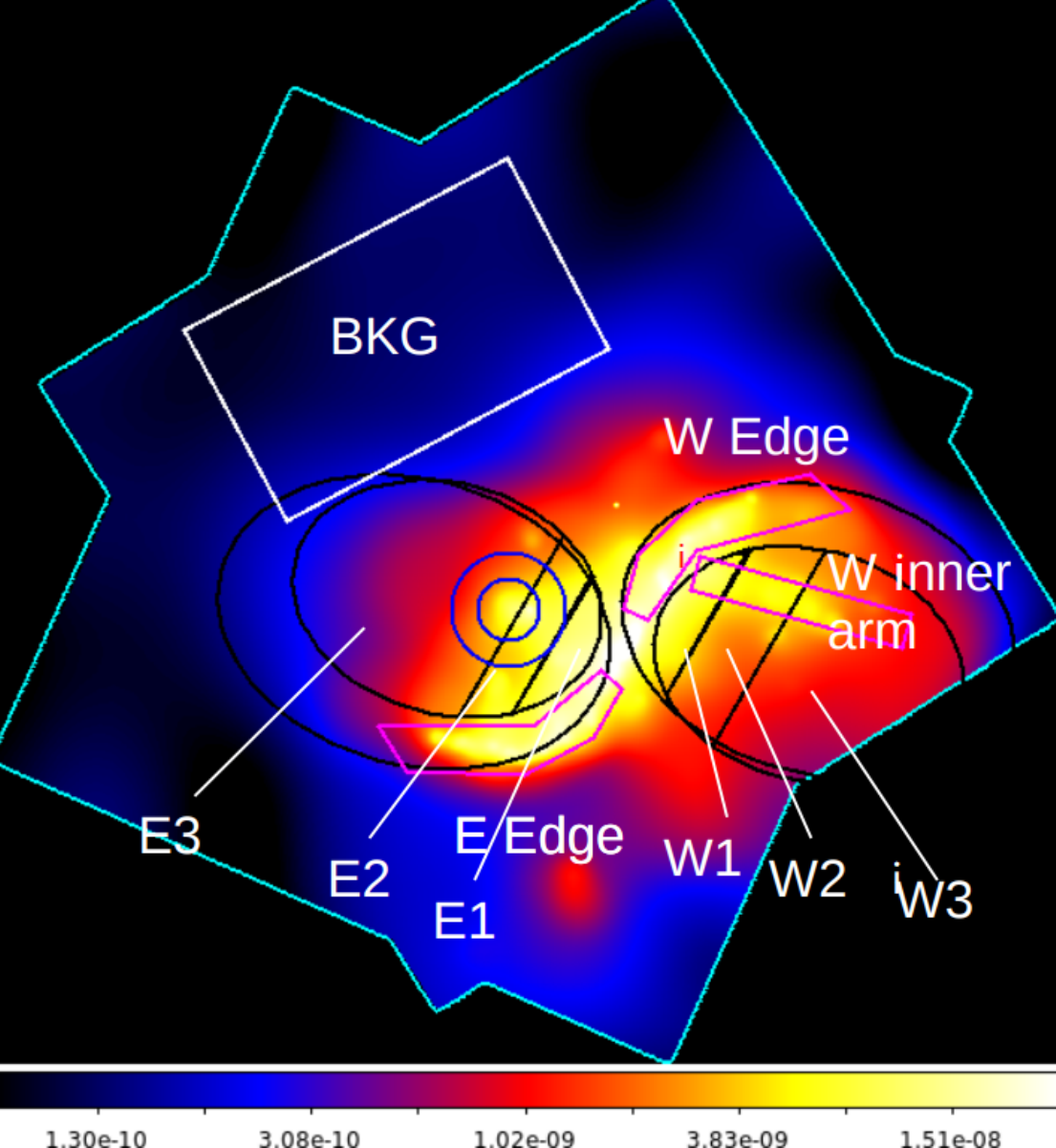}
\caption{Diffuse X-ray emission map of M106. This map is constructed in the 0.45-1 keV band after the removal of detected sources and smoothed with the CIAO routine csmooth with S/N >3. Several spectral extraction regions are outlined: the two large ellipses for the entire E and W bubbles, as well as their (E and W) interiors (represented by the two smaller ellipses) and (E and W) edges, which are the same as in Fig.~\ref{fig:f3}. Interior segments of the bubbles and the background field (BKG) of the bubbles (same as in Fig.~\ref{fig:f1}) are also marked. In addition, the small blue circle inside the E bubble encloses a hot spot with respect to its local background estimated in the annulus defined by the two blue circles. The cyan contour is the same as in Fig.~\ref{fig:f5}.
}
  \label{fig:f6}
\end{figure}

We use the broad-band count and exposure maps to detect discrete sources using the WAVEDETECT tool at scales of 1.0, 1.4, 2.0, 2.8, 4.0, 5.7, and 8.0 pixels. The detected sources are shown in Fig.~\ref{fig:f5}B. The data within 1.2 times the 90\% energy-encircled region (EER) of each source are excluded in our analysis of the diffuse X-ray emission of the galaxy.
For this analysis, we also need to consider the contributions from both non-X-ray events and the local sky X-ray background. The latter component is estimated from a spectrum extracted from a rectangular region northeast of M106, labeled "BKG" in Fig.~\ref{fig:f6} (the same regions used in the other bands), while the former is estimated from the data taken when the telescope was stowed out of the focal plane and under the shield\footnote{http://cxc.harvard.edu/contrib/maxim/stowed/}. We select the stowed data taken on the dates closest to the epoch of the \xs\ observations and reprocess them to match the observations in terms of both roll angles and count rates detected in the 10-12 keV band, where little X-ray contribution is expected. We subtract the resulting non-X-ray event component from the subsequent imaging and spectroscopic analyses. 

\subsubsection{Spatial Analysis}
In addition to mapping the X-ray emission in the different bands, we also produce 1-D plots that allow for a more quantitative assessment of the intensity distributions and a comparison with multi-wavelength data. Specifically, these plots are generated in the three rectangular regions outlined in Fig. ~\ref{fig:f5}. We adaptively divide the parallel region (relative to the galaxy's minor axis) into vertical slices roughly from east to west, each containing a similar number of counts ($\sim 100$). 
Similar divisions are made in the two vertical regions. For comparison, we use the same divisions to calculate the radio intensity distributions with the \rins\ data. 

\subsubsection{Spectral Analysis}\label{s:spec}

Our spectral analysis of the diffuse X-ray emission uses XSPEC, which is part of the HEASOFT v6.31 software suite. We first characterize the local sky X-ray background spectrum after subtracting the non-X-ray contribution. While the procedure is detailed in Appendix \ref{a:spec_bkg}, we find the best-fit model characterization satisfactory and thus use it to predict the X-ray background contributions in different on-source regions (Fig.~\ref{fig:f6}), taking into account the differences in sky coverage and effective exposure. The non-X-ray and local sky X-ray contributions are combined and then subtracted from an on-source spectrum before further analysis. 

Our on-source spectral analysis aims to provide a simple characterization of the thermal and chemical properties of the diffuse hot plasma associated with the bipolar superbubble structure of \xs. 
We extract spectral data not only from the two entire bubbles but also separately from their interiors and from the bright edge regions, as shown in Fig. ~\ref{fig:f6}. To allow differential spectral analysis of the bubble interiors, we further divide them into segments, E1-E3 and W1-W3 (Fig. ~\ref{fig:f6}). The sizes of these segments are adjusted so that they contain a similar number of counts 
($\sim 1700$ for the eastern division and 1250 for the western division) in the 0.45-1 keV band. The spectra from the entire bubbles or the interiors are adaptively grouped to ensure an S/N > 3 per bin, where S is the net number of counts after background subtraction, while N is the Poisson error of the total on-source counts of the bin. The other spectra (from the individual segments or edges) are binned to have S/N > 2. 

\begin{figure}
  { \centering
\includegraphics[width=\columnwidth]{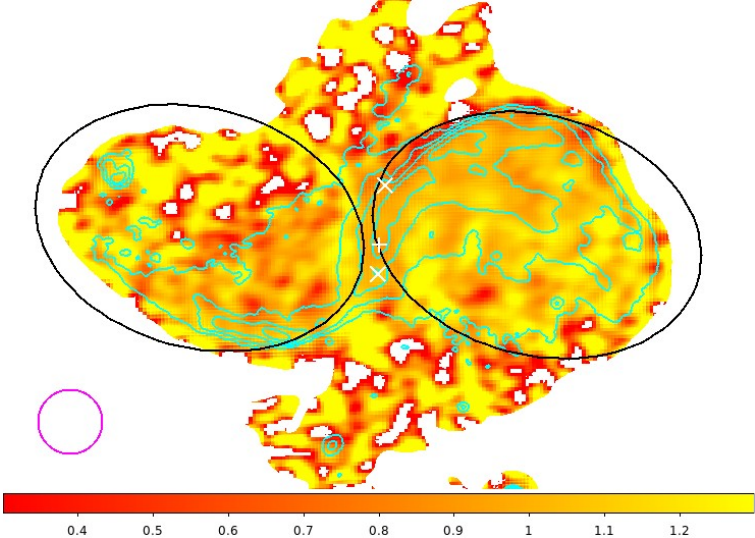}}
 \caption{\rins\ radio spectral index map of M106, compared with the \rins\ 144 MHz intensity illustrated by the cyan contours at (10, 20, 30, 60, 120, 150, 220) ${\rm~mJy~beam^{-1}}$. The two large ellipses mark the bubbles as in Fig.~\ref{fig:f6}. The magenta circle in the lower left corner has a 1 kpc radius at the distance of the galaxy and the resolution of the image is $15^{\prime\prime}$.}
  \label{fig:f7}
\end{figure}
The limited count statistics and spectral resolution of the spectral data allow only relatively simple modeling. However, we find that an optically thin one-temperature (1-T) thermal plasma (APEC) is far from being statistically acceptable (e.g., Table~\ref{t:t3}). Studies based on hydrodynamical simulations \citep[e.g.,][and discussion in \S~\ref{ss:dis_comp_sim}]{Pillepich2021} suggest that the temperature within such bubbles is widely distributed, e.g., ranging from several $10^6$ K (typically found at the outer boundary) to $\gtrsim 10^7$ K (in the interior near galactic centers). In comparison, the variation of the plasma thermal pressure is relatively small (e.g. typically < a factor of 2, probably except for regions near galactic disks), apparently due to the short dynamic (or sound-crossing) time scale of the hot plasma. The X-ray emission measure (EM) of the plasma is thus approximately $\propto n_e n_H \propto T^{-2}$ (where the electron density $n_e \sim n_H$; see further discussion in \S~\ref{ss:dis_thermal}). Accordingly, we adopt a simple plasma model with a lognormal temperature distribution VLNTD \citep{Cheng2021,Wang2021}, which has the key parameters as $\mathrm{\Bar{x}=ln(\Bar{T})}$ and $\sigma_x$ -- the emission-weighted mean and dispersion of the temperature in logarithmic form. Other parameters such as metal abundances and normalization are the same as in the VAPEC model. 
The suitability of using the lognormal temperature distribution to describe the thermal properties of the hot CGM has also been recently demonstrated by \citet{Vijayan2022}.
In any case, we find empirically that this plasma model plus a foreground absorption [or TBABS(VLNTD)] gives a reasonably good characterization for most of our spectra.
\begin{figure}
  \includegraphics[width=\columnwidth]{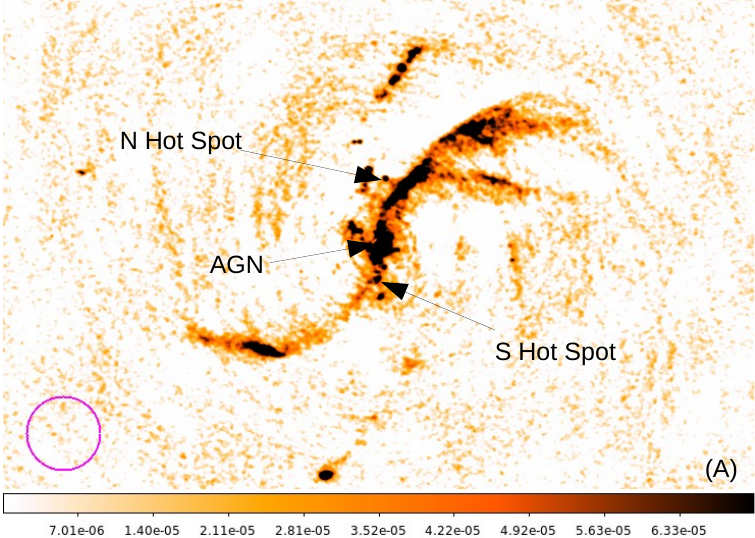}
  \includegraphics[width=\columnwidth]{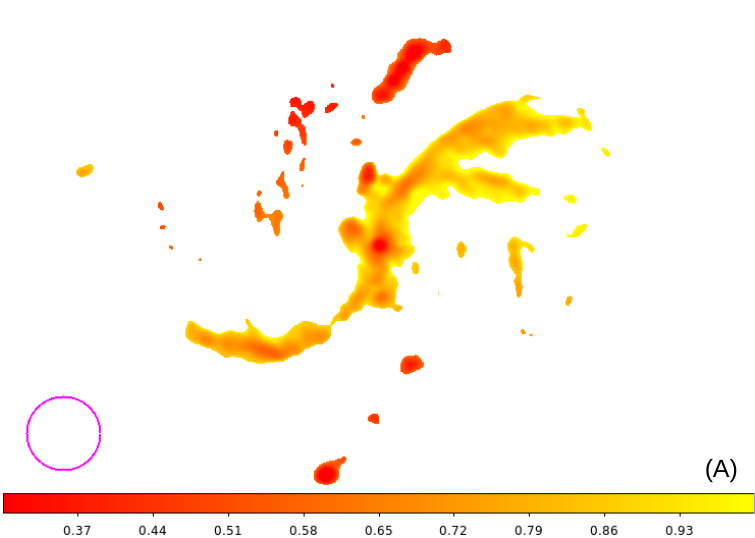}
\caption{(A) VLA 8.44 GHz intensity map (in units of Jy~beam$^{-1}$; \citet{Krause2004}) and (B) spectral index map constructed from this map and the \rins\ 144 MHz map. 
}
  \label{fig:f8}
\end{figure}
\section{Results}\label{s:res}
The presence of a prominent large-scale bipolar superbubble structure with enhanced diffuse radio emission is evident in Fig.~\ref{fig:f1}. This structure is nearly perpendicular to the major axis of the galactic disk of \xs\ (including the two grand-design spiral arms, clearly visible in the 144 MHz band; see also Fig.~\ref{fig:f3}). Fig.~\ref{fig:f2} presents a simplistic illustration of the major components of \xs\ that are most relevant here. The physical link of the bubbles to the disk and/or nucleus of the galaxy is not immediately clear in the available data and will be discussed in \S~\ref{ss:dis_comp_sim}. Fig. \ref{fig:f3} shows a close-up of the structure and a comparison with the multi-wavelength data of the galaxy. The diffuse radio emission associated with the structure is bounded on its southeastern and northwestern sides by the previously known east and west anomalous arms (or E and W edges in Fig.~\ref{fig:f2}), which are offset from the normal spiral arms seen in the HST or GALEX FUV images of the galaxy (Fig.\ref{fig:f3} C-D). Therefore, these two anomalous arms appear to represent parts of the outer boundaries of the bipolar super-bubble structure of the diffuse radio emission. There is another arm-like feature that we indicate as W inner arm (Figs. \ref{fig:f2} and \ref{fig:f6}) and whose nature is not clear. It could be a bifurcated branch of the W edge or a separate outflow probably driven by the N hot spot (apparently seen in high-resolution radio images \citep[e.g.][]{Krause2004}. 
The total fluxes are 75/26 Jy at 54/144 MHz within the fields of the two ellipses and 33/13 Jy within the two bright edges (outlined regions in Fig.~\ref{fig:f3}).
\begin{figure}
  \includegraphics[width=\columnwidth]{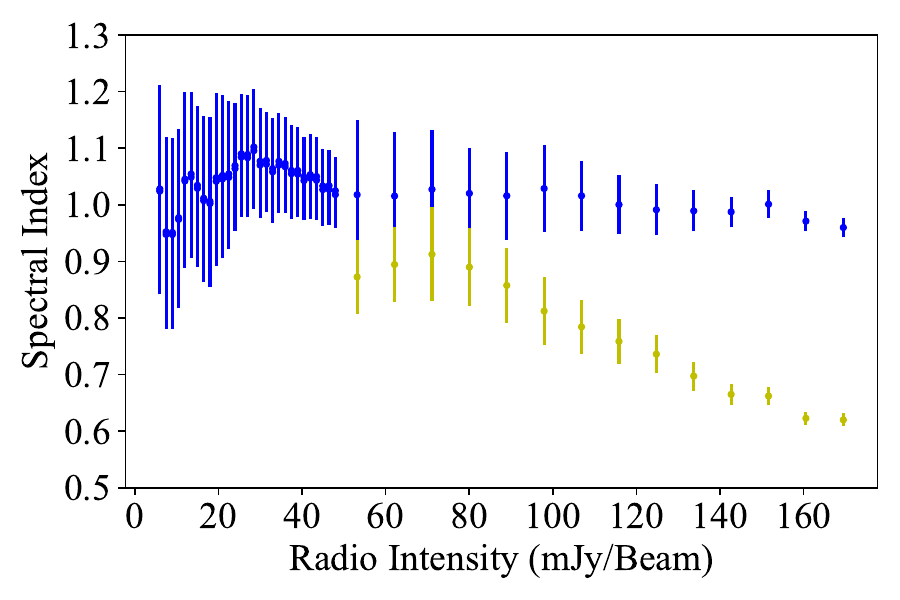}
  \caption{Radio spectral index vs. original \rins\ 144~MHz intensity of the \xs\ structure: the blue index points are calculated with the \rins\ 54 MHz and 144 MHz data over the entire bubble regions at the 15" resolution, while the yellow ones are with the \rins\ 144 MHz and VLA 8.44 GHz data over the anomalous arms at the 6" resolution (Fig.~\ref{fig:f8}). The error bars represent the RMS of the spectral index values within each intensity bin.
  }
  \label{fig:f9}
\end{figure}

Figure \ref{fig:f7} shows the radio spectral index map constructed from the \rins\ data. The spectral index ($\alpha \approx 1$) clearly indicates the synchrotron nature of the radio emission and shows only a small region-to-region variation of typically $\lesssim 15\%$. 
However, this result should be taken with caution due to the limited spatial resolution ($15^{\prime\prime}$) of the spectral index map. It is possible that variations are present at smaller spatial scales, especially in the regions of the anomalous arms, whose widths are indeed below the spatial resolution of this map.

To further explore the properties of the anomalous arms, we re-examine the VLA 8.44 GHz data (Fig.~\ref{fig:f8}A) described by \citet{Krause2004}, together with the \rins\ image. 
The VLA data, obtained with the C-array configuration, has a resolution of 2".2 $\times$ 2".4 and the RMS of $\sim 8~\mu{\rm Jy~beam^{-1}}$. This high-resolution radio image shows a morphological indication for the possible connection of the W inner arm to the N hot spot (Fig.~\ref{fig:f2}). However, the complexity of the emission in the region makes it difficult to formulate any conclusive assessment of this scenario. Alternatively, the inner arm could simply be a bifurcation of the W edge, as proposed by a detailed study of the jets and anomalous arms \citep{Cecil2000}. We construct a spectral index map, using the VLA 8.44 GHz data and the \rins\ 144 MHz image. To do so, we convolve the VLA data with a Gaussian to match the resolution of the \rins\ image and apply the same S/N $> 3$ threshold as used for the \rins\ index map construction. Fig.~\ref{fig:f8}B shows the result. The spectral index of the anomalous arms ($\alpha \approx 0.8$) is broadly consistent with the values obtained from using only VLA data at 1.46, 1.49, 4.88, and 5 GHz \citep{Hummel1989,Hyman2001}, but is systematically larger than that of the normal spiral arms ($\alpha \approx 0.45$, presumably due to the free-free emission contribution from HII regions). 

The spectral index obtained from the LOFAR and VLA data is strongly anti-correlated with the 8.44 GHz intensity along the anomalous arms (Fig. ~\ref{fig:f9}). We may assume that the index ($\sim 0.6$) at the highest intensity end is due to the synchrotron emission of the cosmic ray electrons before any significant cooling.
The spectral index obtained with the \rins\ 54-144 MHz data shows a similar but milder anti-correlation at the 144 MHz intensity $\gtrsim 25 {\rm~mJy~beam^{-1}}$. The flattening of the anti-correlation at the intensity $\gtrsim 45 {\rm~mJy~beam^{-1}}$ is probably due to the limited spatial resolution of the \rins\ 54 MHz data. The 15" beam of the \rins\ index map, which is considerably larger than the typical width of the anomalous arms ($\sim 7^{\prime\prime}$), is significantly contaminated by diffuse radio emission with a steeper spectrum with a characteristic index of probably $\sim 1.1$, as may be expected from synchrotron cooling. If we assume that the LOFAR-LOFAR spectral index is the same as the LOFAR-VLA index we can estimate that the contamination is about 40\%, as estimated from the high-resolution VLA data at 8.44 GHz, and increases with the decreasing frequency because of its steeper spectrum. Largely as a result of the contamination, the \rins\ 54-144 MHz index becomes saturated at $\sim 1$ along the anomalous arms.
\begin{figure}
  \includegraphics[width=\columnwidth]{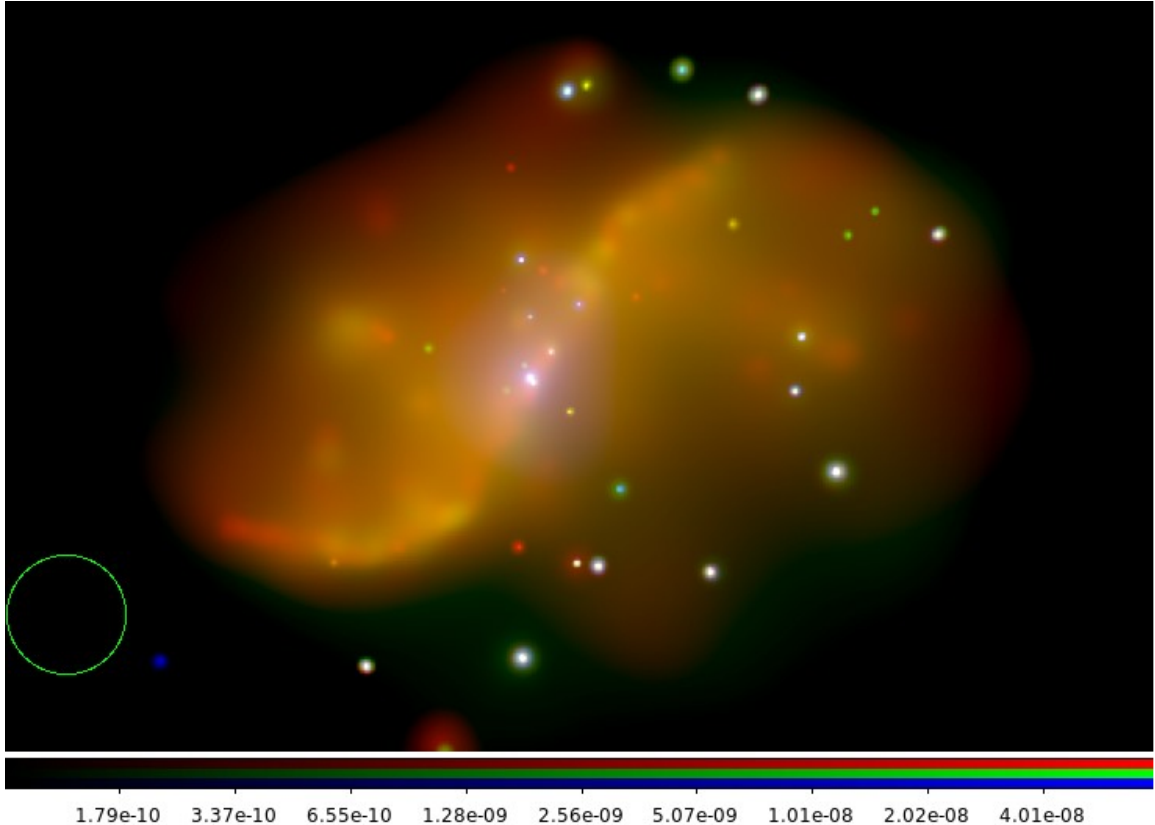}
  \caption{3-color composite of the intensity images of \xs\ in the 0.45-1 keV (red), 1-2 keV (green), and 2-7 keV (blue) bands. These images have been smoothed with the CIAO CSMOOTH routine to achieve S/N > 3 this is adaptive smoothing so the resolution is not given as other Gaussian smoothing. The radius of the green circle at the lower left corner illustrates the 1~kpc scale. 
  }
  \label{fig:f10}
\end{figure}
\begin{figure}
  \includegraphics[width=\columnwidth]{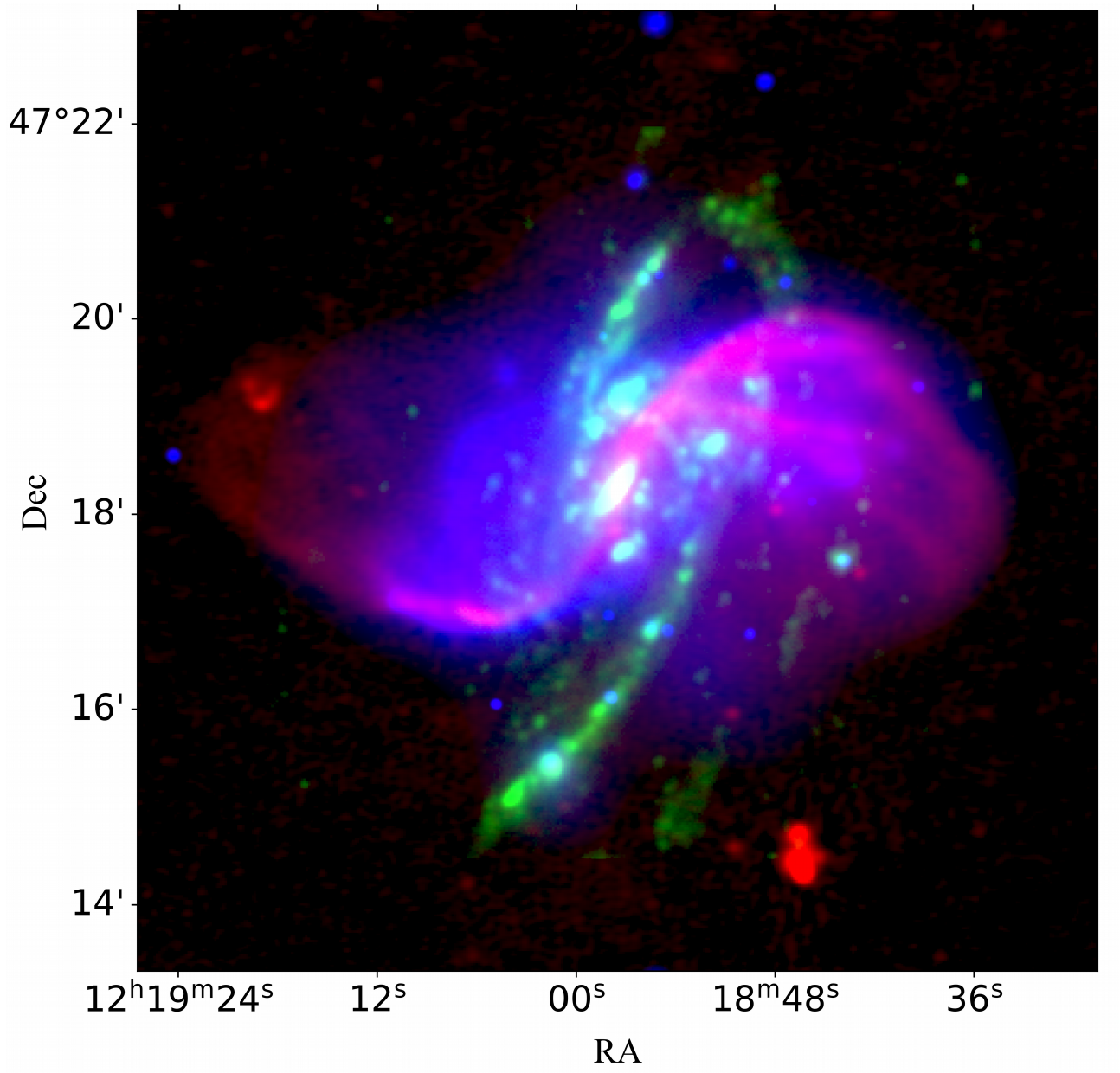}
 \caption{ 3-color composite of the intensity images of \xs\ in the \rins\ 144 MHz (red), GALEX FUV (green), and Chandra 0.45-1 keV (blue) bands.}
  \label{fig:f11}
\end{figure}

The radio bubbles are also visible in the X-ray images (Figs. \ref{fig:f6}, \ref{fig:f10}, and \ref{fig:f11}). The X-ray enhancement, most pronounced in the 0.45-1 and 1-2 keV bands, traces hot plasma emission, in contrast to the 2-7 keV band, which is dominated by point-like sources. In Fig. ~\ref{fig:f10}, the large-scale orange-colored diffuse emission away from the central galactic disk has an overall morphology similar to the radio bubbles and their rim-brightened edges (see also Fig. ~\ref{fig:f11}), while green or white dots represent point-like sources that are excluded from the analysis of the diffuse X-ray emission (e.g. Fig.~\ref{fig:f6}). 
The similarity between X-rays and radio is most striking for the W bubble (Fig. ~\ref{fig:f11}) with comparable total off-disk extents. For the E bubble, however, the diffuse X-ray emission is enhanced near the galactic disk and drops off steeply beyond about half the \rins\ bubble extent. Part of this enhancement is due to a feature that we refer to as the eastern "hot spot" in Fig.~\ref{fig:f6}. This X-ray feature, however, has no apparent multi-wavelength counterpart. The limited counting statistics of the X-ray data prevent us from a detailed 2-D study of the diffuse X-ray emission substructure. 

\begin{figure}
  \includegraphics[width=\columnwidth]{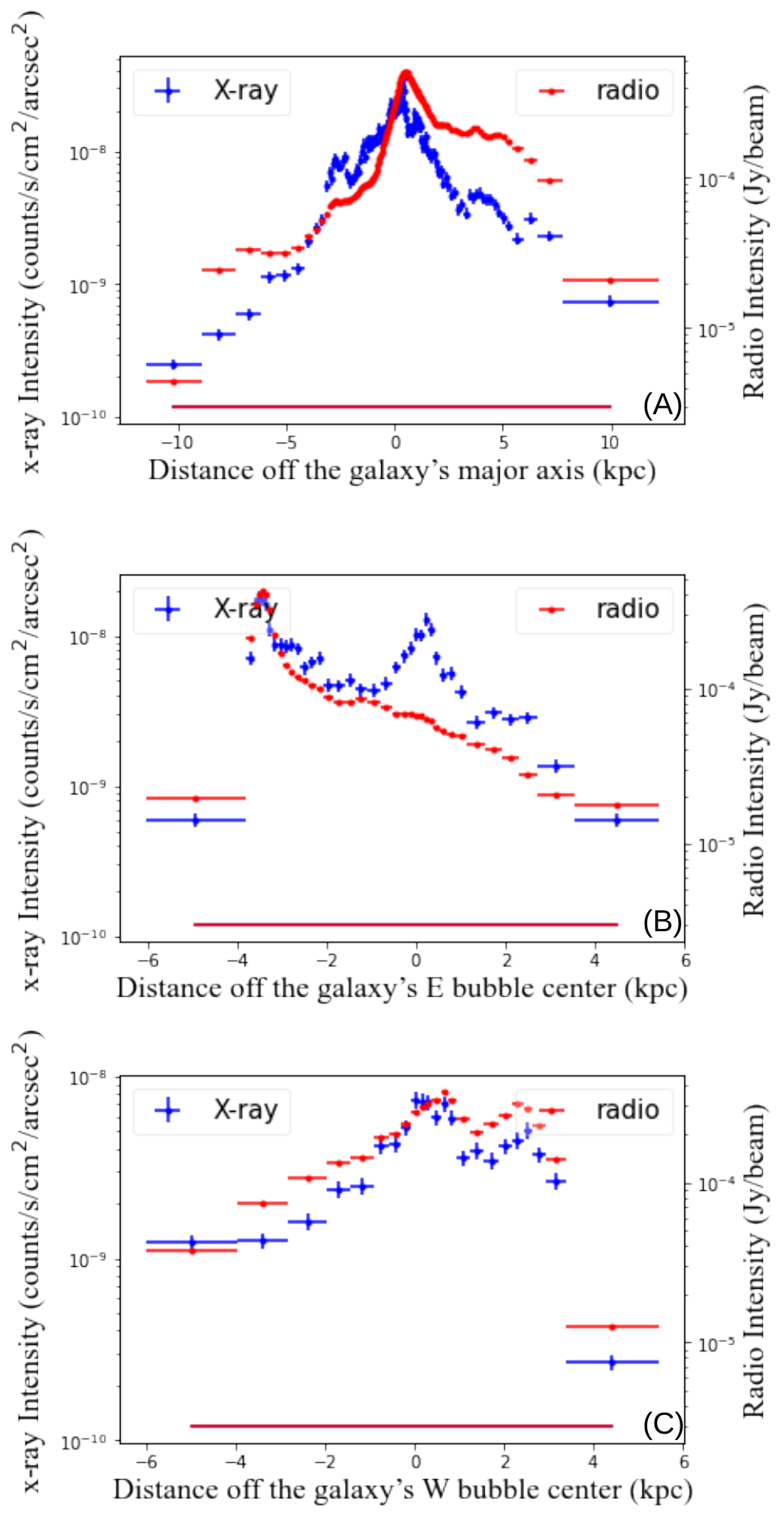}
  \caption{1-D intensity profiles along the cuts shown in Fig.~\ref{fig:f5}: \rins\ 144 MHz (red) and the 0.45-1~keV diffuse emission (blue). The 
  parallel plot (A) has its coordinate centered on the major axis of the galaxy (positive toward the southwest), while the vertical plots (B - east cut; C - west cut) are centered on the ellipse centers of the bubbles (positive toward the northwest). The straight horizontal lines mark the local radio and X-ray background levels. The positive offsets of the data points above the levels are due to the presence of the radio/X-ray-emitting CGM even outside the bubbles.}
  \label{fig:f12}
\end{figure}

Fig.~\ref{fig:f12} presents the intensity distributions across the three rectangular cuts shown in Fig.~\ref{fig:f5}. We use these distributions to examine the overall dimensions of the two bubbles. The (mostly east-west) distribution parallel to the minor axis of the galaxy (Fig.~\ref{fig:f12}A) shows that the radio intensity drops sharply at $\sim 8$~kpc away from the center of \xs, which can be considered as the total extent of the bubbles above and below the galactic disk; the projection correction for the disk inclination of 71$^\circ$ is only 6\%, well within our estimation uncertainty. Interestingly, Fig.~\ref{fig:f12}A shows that the X-ray intensity generally decreases faster than the radio emission with the distance from the major axis of the galaxy. Within 3 kpc of the major axis of the disk, the X-ray intensity is typically higher on the eastern side than on the western side, which is at least partly due to the absorption effect of the galactic disk. In addition, the distribution shows a shoulder at about 6 kpc on both sides. This shoulder may represent an outer shell-like feature of the bipolar structure, but this needs to be confirmed with better X-ray data. The cuts vertical to the minor axis of the galaxy (Fig.~\ref{fig:f12}B-C) show the central X-ray enhancements as well as a separate peak about 4 kpc south of the E bubble center or 3 kpc north of the W bubble center, corresponding to their rim-brightened edges. The edge brightening is not apparent on the other sides of the bubbles; nevertheless, a steep radio/X-ray intensity drop is seen at $\sim 2-3$~kpc off-center distances, north and south for the E and W bubbles, respectively. The enhancements near the minor axis of the galaxy are largely due to discrete features: the X-ray hot spot in the east and the W inner arm in the west (Figs.~\ref{fig:f2}, \ref{fig:f6} and \ref{fig:f11}).

\begin{figure}
   \includegraphics[width=\columnwidth]{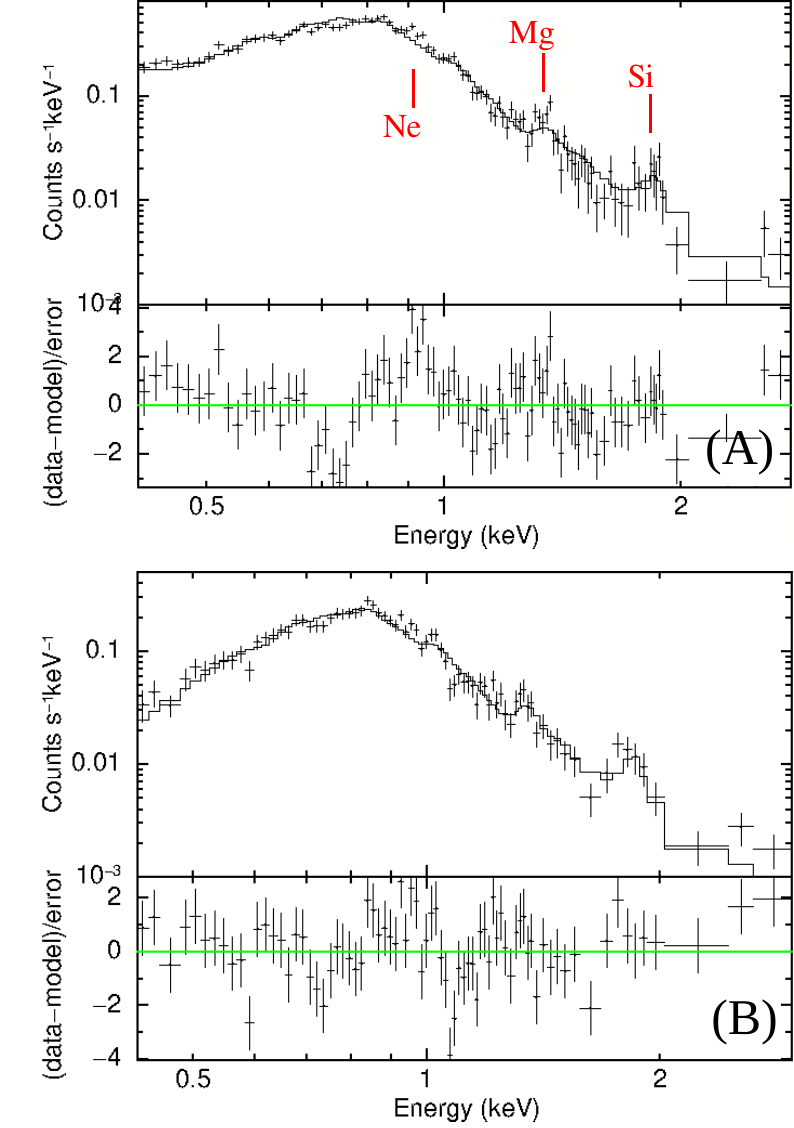}
    \caption{X-ray spectra of the E and W bubbles (A and B panels; Fig. ~\ref{fig:f6}), together with the best-fit TBABS(VLNTD) models (Table~\ref{t:t4}). These fits are not satisfactory, especially for some of the prominent He-$\alpha$ transitions expected for the thermal emission, which are marked in (A) for reference.}
    \label{fig:f13}
\end{figure}
\begin{figure}
    \includegraphics[width=\columnwidth]{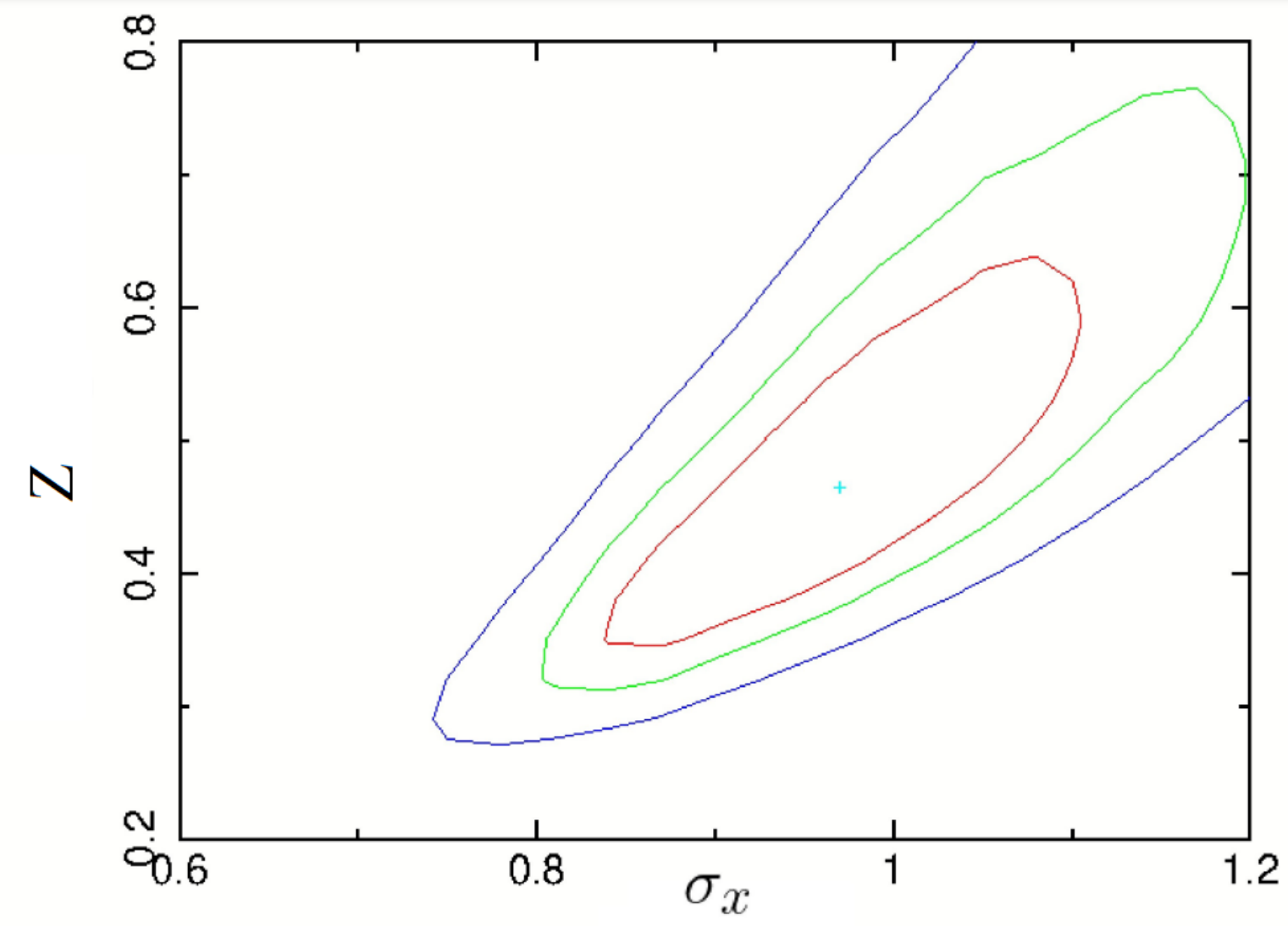}
    \caption{Illustration of the anti-correlation between the metal abundance $Z/Z_\odot$ and the lognormal temperature dispersion $\sigma_x$ in the TBABS(VLNTD) model fit of the  E bubble spectrum (Fig.~\ref{fig:f13}). The confidence contours are at 68.3\%, 95.4\%, and 99.7\% around the best fit, marked as the plus sign (Table~\ref{t:t4}).}
    \label{fig:f14}
\end{figure}

\begin{figure}
    \includegraphics[width=\columnwidth]{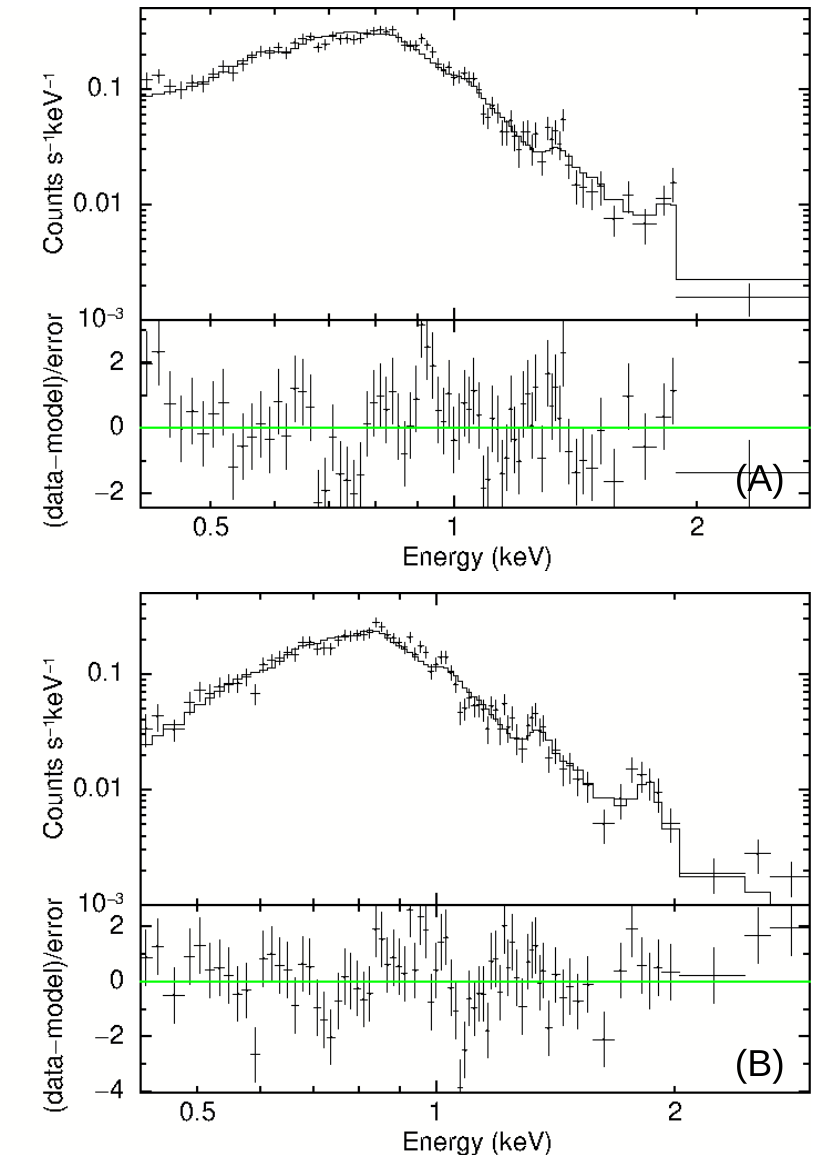}
    \caption{X-ray spectra of the E and W interiors (A and B panels; Fig. ~\ref{fig:f6}), together with the best-fit TBABS(VLNTD) models (Table~\ref{t:t4}).}
    \label{fig:f15}
\end{figure}
\begin{figure}
    \includegraphics[width=\columnwidth]{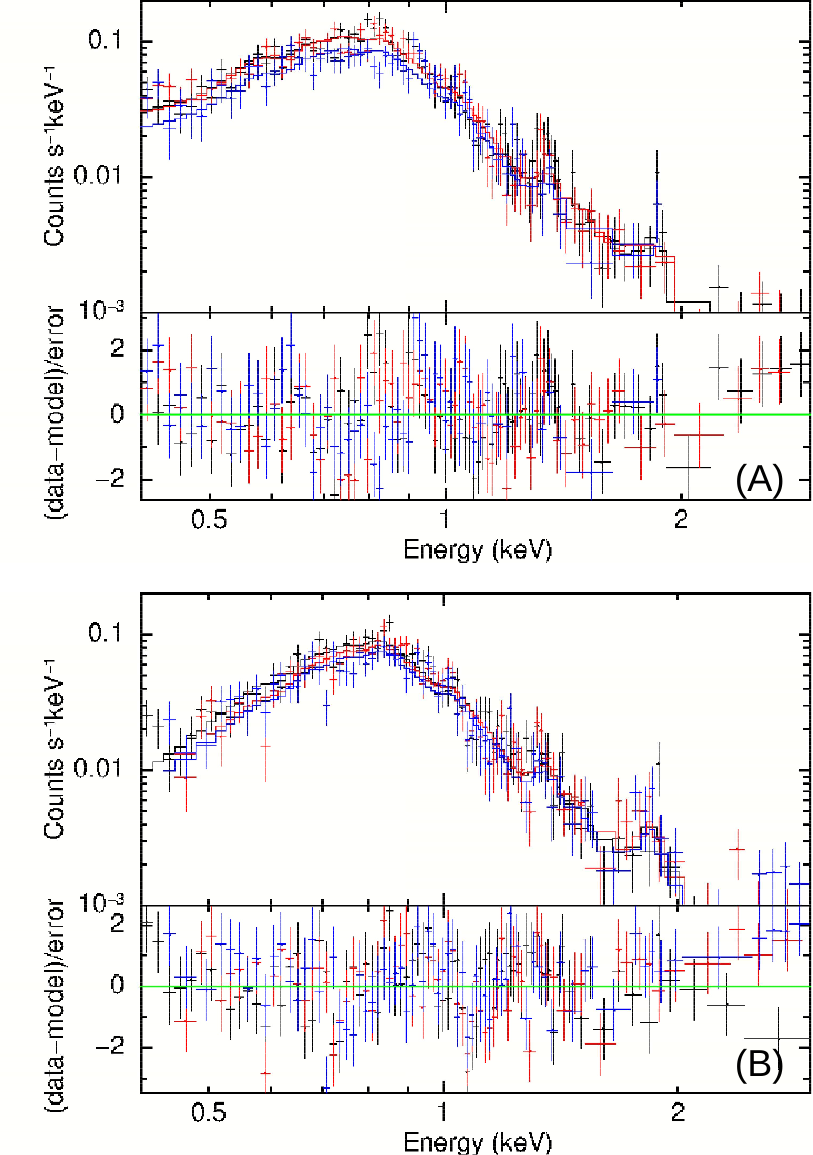}
 \caption{The joint TBABS(VLNTD) model fits of the X-ray spectra extracted from the segments of the E and W interiors (Fig. ~\ref{fig:f6}), separately: (A) the east segment set [E1 (black), E2 (red), and E3 (blue)] and (B) the west segment set [W1 (black), W2 (red), and W3 (blue). For each segment set, both the fixed absorption and metal abundance, as well as the jointly fitted temperature dispersion and individually fitted mean temperatures are given in Table~\ref{t:t5} (see text for details).}
    \label{fig:f16}
\end{figure}
Fig.~\ref{fig:f13} shows the X-ray spectra of the two bubbles. Due to the overall steep shape, as well as apparent emission line features that can be identified as being due to transitions such as Ne, Mg, and Si He-$\alpha$, the spectra must be primarily thermal. Table~\ref{t:t3} presents our spectral fit results based on the 1- or 2-T APEC plasma modeling, chiefly for comparison with previous similar studies (see \S~\ref{ss:dis_comp_obs}). We present our results mainly from the TBABS(VLNTD) modeling (Tables~\ref{t:t4}), which is more physically realistic.  
Our spectral analysis shows that the foreground absorption is consistent with the hypothesis that the E bubble is on the near side of the galactic disk (e.g., Table~\ref{t:t3}; Fig.~\ref{fig:f2}). Therefore, we fix the absorption to the known Galactic HI column density $N_{\rm H,G}$ (Table~\ref{t:t1}). In contrast, the fitted N$_{\rm H} \approx 2.3 \times 10^{21} {\rm~cm^{-2}}$ for the W bubble is larger than N$_{\rm H,G}$, consistent with its location on the far side of the disk. We find that both the mean temperature and the X-ray luminosity of the plasma in the W bubble are consistently higher than in the E bubble. 
The different spectral models give very different values for the metal abundance ($Z$) of the plasma (Table~\ref{t:t4}). The value increases from 1-T APEC, to 2-T APEC, and to VLNTD, reflecting their increasing proximity to the real temperature distribution of the plasma. However, in the VLNTD model, $Z$ is strongly correlated with $\sigma_x$ (Fig.~\ref{fig:f14}), leading to the larger fitting errors reported for these parameters in Table~\ref{t:t4}. 
Although the TBABS(VLNTD) modeling of the entire bubbles gives a reasonable characterization of their overall X-ray spectral shapes, the fits have large reduced-$\chi^2/$dof values. 

We thus further present the results of the TBABS(VLNTD)  model fitting to spectra extracted from the sub-regions of the bubbles. Fig.~\ref{fig:f15} shows the fits to the data from the E and W bubble interiors, which cannot be rejected at the statistical confidence  $\gtrsim 3\sigma$ (Fig.~\ref{fig:f6}; Table~\ref{t:t4}). Even better fits are obtained for the spectra from the individual segments of the bubble interiors (Fig.~\ref{fig:f6}; Table~\ref{t:t4}). The best-fit mean temperature seems to decrease with increasing distance from the galaxy's major axis, i.e. from E1 to E3 and from W1 to W3. However, the temperature dispersion ($\sigma_x$) shows an opposite trend, increasing with distance, because the two parameters are statistically anti-correlated in the spectral fits. To minimize this degeneracy effect and to check how the mean temperature might vary between the segments, we perform a joint fit of the E1-E3 and W1-W3 spectra (Fig.~\ref{fig:f16}) with the common fitting temperature dispersion $\sigma_x$ and with both the metal abundance and the absorption column fixed at the best-fit values for the E and W interiors (Table~\ref{t:t4}). The fitted parameters are listed in Table~\ref{t:t5}. While the quality of the fit does not change much (as judged by the $\chi^2$/dof values), the trend of decreasing mean temperature with distance disappears for both bubbles. Interestingly, the E bubble has both a higher mean temperature and $\sigma_x$ than the W bubble when the abundance and absorption column are fixed. Table~\ref{t:t4} also includes the results for the E hot spot and the W inner arm based on similar spectral fits. We find that 
the E hot spot is indeed hotter than its surroundings, whereas the inner arm appears slightly cooler compared to other parts of the W interior, but is consistent with the W edge. The inner arm has a  value of $N_H$ (if allowed to be fitted) consistent with being behind the galactic disk.

\begin{table*}
	\centering
	\caption{1-T and 2-T plasma model fit results}
	\label{t:t3}
	\begin{tabular}{lccccccccccr} 
		\hline
		Region&area&nH&$k_B T_1$&$k_B T_2$&Z&$K_{th1}$&$K_{th2}$&$\chi^{2}$/{\rm dof}&$f_x$&$L_x$\\
  &arcmin$^2$&10$^{20}$cm$^{-2}$&keV&keV&Z$_\odot$&10$^{-3}$cm$^{-5}$&10$^{-3}$cm$^{-5}$&&10$^{-13}$erg cm$^{-2}$s$^{-1}$&10$^{39}$erg s$^{-1}$\\
		\hline
		E Bubble&12.14&0$^{<8.22}$&0.53$_{-0.01}^{+0.03}$& &0.07$_{-0.01}^{+0.01}$&1.65$_{-0.10}^{+0.08}$& &422/92&5.69&6.30\\
        - &- &0$^{<0.89}$&0.24$_{-0.01}^{+0.02}$& 0.76$_{-0.02}^{+0.03}$&0.26$_{-0.04}^{+0.05}$&0.58$_{-0.07}^{+0.10}$& 0.46$_{-0.07}^{+0.08}$&110/90&6.18&6.48\\
		W Bubble&12.14&1.58$_{-1.09}^{+1.31}$&0.69$_{-0.02}^{+0.02}$& &0.11$_{-0.01}^{+0.01}$&1.39$_{-0.13}^{+0.13}$& &344/127&5.80&7.20\\
        - &- &10.4$_{-0.21}^{+0.23}$&0.25$_{-0.02}^{+0.03}$& 0.76$_{-0.03}^{+0.04}$&0.18$_{-0.03}^{+0.03}$&1.28$_{-0.31}^{+0.37}$& 0.98$_{-0.15}^{+0.14}$&188/125&6.06&11.1\\
		\hline
	\end{tabular}\\
	\raggedright
       Note: Listed parameters of the best-fit TBABS(APEC) (1st row) or TBABS(APEC$_1$+APEC$_2$) (2nd row) for each bubble: $T_1$ - the temperature of the APEC or APEC$_1$ plasma; $T_2$ - the temperature of the APEC$_2$ plasma; $K_{th1}$ and $K_{th2}$ - the corresponding normalizations of the two plasma components; $\chi^{2}$/{\rm dof} where dof is the degree of freedom of each fit. 
Also listed are the derived parameters: $f_x$ - the absorbed flux in the 0.45-1 keV range and $L_x$ - the (unabsorbed) 0.1-10 keV luminosity. All error bars are measured at the 90\% confidence level.
\end{table*}
\begin{table*}
	\centering
	\caption{lognormal temperature plasma model fit results}
	\label{t:t4}
	\begin{tabular}{lccccccccr} 
		\hline
		Region&Area&$N_H$&$k_B \bar{T}$&$\sigma_x$&Z&$K_{th}$&$\chi^{2}$/{\rm dof}&$f_x$&$L_x$\\
  &arcmin$^2$&10$^{20}$cm$^{-2}$&keV&&Z$_\odot$&10$^{-3}$cm$^{-5}$&&10$^{-13}$erg cm$^{-2}$s$^{-1}$&10$^{39}$erg s$^{-1}$\\
		\hline
		E Bubble &12.14&0$^{<1.14}$&0.31$_{-0.01}^{+0.03}$& 0.97$_{-0.06}^{+0.15}$&0.46$_{-0.13}^{+0.18}$&1.29$_{-0.23}^{+0.28}$& 165/91&6.02&6.52\\
        E Interior&7.70&4.21(fix)&0.25$_{-0.08}^{+0.06}$&1.02$_{-0.20}^{+0.31}$&0.42$_{-0.15}^{+0.31}$&1.16$_{-0.16}^{+0.29}$&101/76&3.52&4.72\\
		E1 &0.62&4.21(fix)&0.27$_{-0.06}^{+0.05}$&0.89$_{-0.15}^{+0.20}$&0.42(fix)&0.36$_{-0.05}^{+0.08}$&88/82&1.22&1.61\\
		E2 &1.18&4.21(fix)&0.23$_{-0.08}^{+0.06}$&1.03$_{-0.17}^{+0.24}$&0.42(fix)&0.41$_{-0.06}^{+0.13}$&77/73&1.20&1.63\\
		E3 &5.90&4.21(fix)&0.15$_{-0.11}^{+0.19}$&1.43$_{-0.14}^{+0.21}$&0.42(fix)&0.47$_{-0.23}^{+0.24}$&70/59&0.99&1.45\\
		E Edge &1.34&4.21(fix)&0.21$_{-0.05}^{+0.04}$&1.01$_{-0.13}^{+0.16}$&0.42(fix)&0.81$_{-0.11}^{+0.17}$&120/85&2.31&3.12\\
        E hot spot&0.42&4.21(fix)&0.35$_{-0.35}^{+0.10}$&0.97$_{-0.19}^{+0.32}$&0.43$_{-0.20}^{+0.71}$&0.14$_{-0.07}^{+0.07}$&83/69&0.54&0.74\\
        W Bubble &12.13&6.31$_{-2.24}^{+4.72}$&0.41$_{-013}^{+0.08}$& 0.96$_{-0.15}^{+0.18}$&0.39$_{-0.11}^{+0.12}$&1.86$_{-0.39}^{+1.13}$& 206/126&5.89&9.25\\
		W Interior& 6.38&23.2$_{-5.4}^{+6.4}$&0.13$_{-0.13}^{+0.20}$&1.12$_{-0.05}^{+0.06}$&0.76$_{-0.27}^{+0.71}$&1.73$_{-0.80}^{+0.85}$&112/74&2.16&7.7\\
		W1 &0.79&23.2(fix)&0.19$_{-0.08}^{+0.05}$&0.85$_{-0.16}^{+0.27}$&0.76(fix)&0.52$_{-0.11}^{+0.33}$&85/77&0.87&3.03\\
		W2 &1.74&23.2(fix)&0.18$_{-0.09}^{+0.09}$&0.96$_{-0.25}^{+0.34}$&0.76(fix)&0.47$_{-0.15}^{+0.33}$&86/70&0.77&2.67\\
		W3 &4.08&23.2(fix)&0.14$_{-0.14}^{+0.10}$&1.12$_{-0.10}^{+0.12}$&0.76(fix)&0.53$_{-0.03}^{+0.03}$&86/69&0.70&2.50\\
		W Edge &1.36&23.2(fix)&0.13$_{-0.05}^{+0.05}$&0.93$_{-0.04}^{+0.04}$&0.76(fix)&1.88$_{-0.07}^{+0.07}$&165/90&2.21&8.19\\
        W inner arm&1.00&23.2(fix)&0.14$_{-0.08}^{+0.14}$&1.01$_{-0.53}^{+0.25}$&0.76(fix)&0.57$_{-0.19}^{+0.09}$&80/71&0.75&1.8\\
		\hline
	\end{tabular}\\
	\raggedright
       Note: Same as the caption to Table~\ref{t:t3}, but for the TBABS(VLNTD) model in which the listed parameters are $\bar{T}$ - the mean temperature; $\sigma_x$  - the dispersion of the temperature in logarithm, and $K_{th}$ - the normalization of the plasma \citep{Cheng2021,Wang2021}. 
\end{table*}

\begin{table}
	\centering
	\tabcolsep=0.2cm  
	\caption{Joint fit results of the VLNTD model to interior segment spectra}
	\label{t:t5}
	\begin{tabular}{lcccccccr} 
		\hline
		Region&$k_B \bar{T}$&$\sigma_x$&$K_{th}$&$\chi^{2}$/{\rm dof}&$f_x$&$L_x$\\
		\hline
		E 1 &0.22$_{-0.05}^{+0.04}$&1.04$_{-0.11}^{+0.15}$&0.42$_{-0.05}^{+0.08}$&90/85&1.20&1.65\\
		E 2 &0.23$_{-0.05}^{+0.04}$&-&0.41$_{-0.05}^{+0.08}$&77/76&1.20&1.65\\
		E 3 &0.24$_{-0.05}^{+0.07}$&-&0.35$_{-0.05}^{+0.06}$&74/62&1.00&1.37\\
		W 1 &0.17$_{-0.05}^{+0.05}$&0.88$_{-0.12}^{+0.15}$&0.96$_{-0.20}^{+0.32}$&86/80&0.86&3.13\\
		W 2 &0.20$_{-0.05}^{+0.05}$&-&0.72$_{-0.14}^{+0.26}$&90/73&0.76&2.69\\
		W 3 &0.19$_{-0.05}^{+0.05}$&-&0.71$_{-0.16}^{+0.25}$&88/72&0.71&2.51\\
		\hline
	\end{tabular}
\end{table}

\section{Discussion}\label{s:dis}

The above results now enable us to infer the thermal and nonthermal properties of the superbubble structure in \xs, to make comparisons with those similar features observed in other galaxies, especially \sou\ in our Galaxy, and with the relevant simulations \citep{Pillepich2021}, and to probe their energy sources. Our goal here is to achieve a better understanding of the formation and evolution of the structures and their potential impacts on the host galaxies.

\subsection{Physical properties of the bubbles in M106}
\label{ss:dis_thermal}

We here infer the physical properties of the diffuse hot plasma enclosed in the bubbles. This inference is based on the spectral fitting results listed in Table \ref{t:t5}, as well as the relevant 
 formulae obtained for the VLNTD \citep{Cheng2021,Wang2021}, which include the thermal pressure
\begin{equation}
\begin{split}
P_{th} & =\sqrt{\frac{4\pi D^2\eta^2K_{th}}{10^{-14}V_{t}}} (k_B \bar{T}) e^{\sigma_x^2}\\
 & \approx (1.74 \times 10^{33} {\rm~keV~cm^{-3}}) \sqrt{\frac{K_{th}}{V_t}} (k_B \bar{T})_{\mathrm{keV}}e^{\sigma_x^2},
\end{split}
\label{e:pressure}
\end{equation}
the total thermal energy  $E_{th} =\frac{3}{2} P_{th} V_{t}$, the integrated emission measure 
\begin{equation}
EM  = \Big[\dfrac{P_{th}}{\eta k_B \bar{T}}\Big]^2 V_t e^{-2\sigma_x^2},
\label{e:em}
\end{equation}
and the total mass of the plasma
\begin{equation}
M_{th}  = \dfrac{P_{th}\mu m_{p} V_{t}}{k_B \bar{T}} e^{-\sigma_x^2/2},
\label{e:mass}
\end{equation}
where D is the distance to the galaxy, $\eta$ is 2.1 for typical hot plasma metalicities, $\mu$ is the atomic weight, $m_p$ is the proton mass, and $V_t$ is the fractional volume of the bubble assumed to have an ellipsoidal shape. The entire bubble has a volume of approximately 150 kpc$^3$. 
The inferred parameters are included in Table \ref{t:t6}. Overall, there is a trend of decreasing thermal pressure with the increasing distance from the galactic center for both bubbles.  We estimate the mass of the plasma in both bubbles to be approximately $10^8 M_{\odot}$ and the cooling timescale of the plasma in a region as $t_{c} \sim E_{th}/L_{bol}$, 
where the bolometric luminosity $L_{bol}$ is approximated as the unabsorbed luminosity $L_x$ integrated over the 0.1-10~keV range (Table.~\ref{t:t5}). 
\begin{table*}
	\centering
	\tabcolsep=0.2cm  
	\caption{Inferred plasma parameters in individual regions of the bubbles}
	\label{t:t6}
	\begin{tabular}{lcccccccccr} 
		\hline
		Region&$V_t$ & $EM_{th}$ &$n_e$ &$M_{th}$ &$P_{th}$ &$E_{th}$ &$t_{c}$\\
		&kpc$^3$ &$10^{62}{\rm~cm^{-3}}$&$f_h^{-1/2}{\rm~cm^{-3}}$&$10^{7}f_h^{1/2}\rm~M_\odot$&$f_h^{-1/2}{\rm~keV~cm^{-3}}$&$10^{56} f_h^{1/2} {\rm~erg/s}$&$f_h^{1/2}$\,Gyr\\
		\hline
		E 1 &9 &2.88&0.033&1.6& 0.040&0.28&0.53\\
		E 2 &23 & 2.81&0.020&2.53&0.029&0.47&0.90\\
		E 3 &118 & 2.40&0.008&5.31&0.012&1.02&2.36\\
        E Interior&150&8.09&0.020&9.44&0.027&1.77&1.26\\
		W 1 &9 & 6.59&0.050&2.08&0.038&0.24&0.24\\
		W 2 &45 & 4.94&0.019&4.03&0.017&0.56&0.66\\
		W 3 &96 & 4.87&0.013&5.85&0.011&0.77&0.97\\
        W Interior&150&16.40&0.027&11.96&0.022&1.58&0.62\\
		\hline
	\end{tabular}\\
   \raggedright
       Note: The row of E or W Interior gives the summed values for $V_t$, $EM_{th}$, $M_{th}$, and $E_{th}$ and the averaged values for $n_e$, $P_{th}$, and $t_{c}$ over the interior segments of the respective bubble.
\end{table*}

\subsection{Nonthermal properties of the radio bubbles in M106}\label{ss:dis_nontherm}
Here we first use our measurements of the radio emission to constrain the magnetic field strength in the bubbles, then explore the implications of the observed anticorrelation between the index and the intensity of the radio emission, and finally estimate the potential inverse Compton (IC) scattering contribution of cosmic ray electrons (including positrons) to the diffuse X-ray emission.

We estimate the magnetic field strength,  assuming the equipartition between the field and cosmic ray energy densities and following the equation \citep{Beck2005}:
\begin{equation}
\begin{split}
        B_{\rm eq} = &\{4\pi(2\alpha + 1)(K_0 + 1) I_{\nu} E_{P}^{1-2\alpha}(\nu/2c_1)^{\alpha}\\
        &/[(2a-1)c_2(\alpha)lc_4(i)]\}^{1/(\alpha+3)},\\
\end{split}
\end{equation}
where $E_P$  is the proton rest mass energy, while the constants $c_1$, $c_2$, and $c_4$ can be found in \citet{Beck2005}, while $\alpha \approx 1$, as obtained above, $i \sim 0$ is the inclination of the emission region with respect to the sky plane, and the constant ratio between the number densities of cosmic-ray protons and electrons $K_0 \sim 100$ is assumed, while  $I_\nu$ is the surface brightness (e.g., in units of ${\rm~erg~s^{-1}~cm^{-2}~Hz^{-1}~sr^{-1}}$), which can be converted from our observed intensity at $\nu=144$~MHz, and $l$ is the path length along the line of sight of the radio-emitting region. The equation shows that the distribution of $B_{\rm eq} l^{1/4}$ is just a function of $I_\nu$ (e.g., Fig.~\ref{fig:f3}B). It is clear that $B_{\rm eq}$ cannot be uniform in the bubbles.  $B_{\rm eq}$ is about the smallest at the bubble centers where $l \sim 6$~kpc is the largest (if the ellipsoid approximation is reasonable) while $I_\nu$ is relatively small). For example,  taking  $I_\nu \sim 20{\rm~mJy~beam^{-1}}$ at  the east bubble center, we get $B_{\rm eq} \sim 4\,\mu$G. The largest $B_{\rm eq}$ tends to be at the bubble edges. With  $l \sim 0.3$~kpc, estimated as the full width of the half peak intensity (Fig.~\ref{fig:f12}B), we estimate $B_{\rm eq} \sim 15\,\mu$G at the east edge. 
Similarly,  we find $B_{\rm eq} \sim 3\,\mu$G and $\sim 14\,\mu$G at the west bubble center and edge. The corresponding magnetic field pressure is $\sim (0.6-4) \times 10^{-3} {\rm~keV~cm^{-3}}$, which is substantially smaller than the thermal pressure in the bubbles (e.g., Table~\ref{t:t6}), suggesting that they are primarily driven by the overpressure of the hot plasma. However, the thermal pressure drops steeply with the increasing distance from the major axis of the galaxy (Table~\ref{t:t6}), while the decline of the radio intensity is much slower (e.g., Fig.~\ref{fig:f12}A). Therefore, the magnetic field pressure becomes more important at the far ends of the bubbles. 

What could be the cause of the anti-correlation between the spectral index and the intensity of the radio emission, as shown in Fig.~\ref{fig:f9}? The most natural explanation is synchrotron steepening, as expected when cosmic ray electrons diffuse out of their accelerating regions with higher magnetic fields.  Following \citet{Murphy2009}, we estimate the synchrotron cooling time scale as $t_{\rm syn}\sim {\rm (1.1 \times 10^8~yr)} (B/10\,\mu{\rm G})^{-3/2} \nu_c^{-1/2} \sim (0.6-7) \times 10^8$~yr at a critical frequency of $\nu_c \sim 144$~MHz and for a nominal field strength of $B_{\rm eq}\sim 3-15\,\mu{\rm G}$ in the bubbles. This time scale appears considerably longer than the age of the structure ($\sim 10^7$~yr; see \S~\ref{ss:dis_origin}).  Therefore, the above use of the nominal field strength is problematic. Analysis based on high-resolution VLA radio data indeed shows that along the anomalous arms,  the equal-partition magnetic field is $\sim 310~\mu{\rm G}$ \citep[e.g.][]{Krause2004}, which gives $t_{\rm syn}\sim 6 \times 10^5$~yr at $\nu_c \sim 144$~MHz or $\sim 1 \times 10^6$~yr at $\nu_c \sim 54$~MHz. The actual magnetic field could be even stronger on smaller scales, leading to even smaller $t_{\rm syn}$. 
This localized cooling in and around where the cosmic-ray electrons are injected explains why the index is $\sim 1-1.1$ in the 54-144~MHz range at the LOFAR resolution and does not change significantly across the bulk of the bubbles.

While we have so far assumed that the thermal plasma dominates the X-ray emission observed from the bubbles, it is important to check if the contribution from the IC scattering may be significant. In or near the galactic disk of M106, seed photons of the process are expected to be due mainly to interstellar dust emission, which peaks at $\sim 100\,\mu m$. \citet{Cecil1995} already show that this IC contribution to the soft X-ray emission in the disk is negligible. At larger distances from the disk, the IC of the cosmic microwave background could contribute more to the observed diffuse X-ray emission. The same electrons that upscatter the background to X-rays should also produce the synchrotron emission in the 10s-100s MHz frequency range  \citep{Cecil1995}. Therefore, the \rins\ data are well-suited to constrain this IC contribution, which we estimate, following the equation \citep{Harris1979}:
\begin{equation}
    S_X = \frac{(5.05\times10^4)^\alpha C(\alpha)G(\alpha)(1+z)^{\alpha+3}S_r\nu_r^\alpha}{10^{47}B_{\rm eq}^{\alpha+1}\nu_X^\alpha}.
\end{equation}
Adopting $C(\alpha)=1.2\times10^{31}$ and $G(\alpha) = 0.5$ since $\alpha\approx 1$,  as well as the redshift of the galaxy  $z=0.0016$, the intensity $S_r \sim 7.5 {\rm~Jy~beam^{-1}}$ at the frequency $\nu_r =144$~MHz, $\nu_X =1.2 \times 10^{17} $~Hz at 0.5 keV, and $B_{\rm eq}\sim 3\,\mu$G, we obtain $S_x \approx 3.3\times 10^{-15}\;  \mathrm{erg}\; \mathrm{cm^{-2}s^{-1}}$, which is only $\sim 1\%$ of the total X-ray flux of the bubbles (Table. \ref{t:t4}). Therefore, we conclude that the IC contribution to the X-ray emission is also insignificant.

\subsection{Comparison with other observations}
\label{ss:dis_comp_obs}
We begin by comparing the \xs\ bubbles with \sou\ and Radio Loop I in our Galaxy. A major advantage of observing the bubbles in \xs\ is the minimal foreground confusion and extinction/absorption, especially towards its E bubble, which is in front of the inclined galactic disk. Indeed, our spectral analysis shows that there is little additional absorption towards the bubble beyond the Galactic one, while the absorption towards the W bubble is considerable (N$_H \sim 2\times10^{21}{\rm~cm^{-2}}$), 
as expected from the disk.  
Our X-ray spectral analysis shows a broad temperature distribution of the plasma in the \xs\ structure. Characterized by the lognormal temperature distribution model, the bubble interiors consistently show $\sigma_x \approx 1$ and $\bar{T}\approx 0.2$~keV (Table~\ref{t:t5}). 
Of course, the quoted temperature characterization is model-dependent. 
Thus, it is more appropriate to compare our alternative 1-T and 2-T plasma characterizations (Table~\ref{t:t3}) with results based on similar modeling for bubbles observed in other galaxies. Based on the 1-T plasma modeling, the mean temperatures of $\approx 0.53$ and 0.69~keV for the E and W bubbles are substantially higher than 0.3~keV for the plasma associated with \sou\ \citep[e.g.,][]{Ursino2016,Miller2016,Kataoka2018}.  The thermal energy of the \xs\ structure is $\sim 3 \times 10^{56} {\rm~erg}$ (Table~\ref{t:t6}), which is comparable to the total energy estimated for \sou\ \citep[e.g.,][]{Kataoka2018,Predehl2020}. 

There are notable differences between the radio/X-ray enhanced edges of the \xs\ structure and \sou. The enhanced edges of the \xs\ structure exhibit an S-shape, which can be naturally explained by the tilted jets of the AGN \citep[e.g.,][]{Cecil2000,Wilson2001,Krause2007}. In contrast, \sou\ is mainly enhanced on the Galactic northeast (positive Galactic longitude/latitude) side, which could be due to the ram pressure of a CGM wind in the halo above the Galactic disk from the Galactic northeast \citep{Mou2023}. In principle, the one-sided enhancement observed in \sou\ could also be produced by recent jet heating. The dissipation of a pair of jets does not have to be symmetric but depends to a large extent on the properties of the respective medium. In this scenario, the jet heating in \sou\ may have ceased some time ago, consistent with the weak mean magnetic field strength ($\sim 4\,\mu$G) in the radio Loop I \citep{Mou2022} compared to our estimated values ($\gtrsim 13\,\mu$G) in the \xs\ edges. 

Now let us compare the \xs\ structure with the bubble northwest of the NGC~4438 nucleus. For this NGC~4438 nuclear bubble (extending $\sim 0.3$~kpc), both 2-T APEC (with $T_{1}= 0.27$~keV and $T_{2}= 1.2$~keV) and 1-T APEC ($T = 0.90$~keV) + power law have been considered, although the latter model is statistically favored \citep{Li2022}.  The 2-T plasma characterization suggests that the plasma in the NGC~4438 bubble is considerably hotter than in the \xs\ E and W bubbles ($T_{1}= 0.24$~keV and $T_{2}= 0.76$~keV).  Incidentally, there is an X-ray faint counter-nuclear bubble of a similar extent on the other side of the galactic disk of NGC~4438. The X-ray faintness of this bubble is largely due to the absorption of the inclined disk - a more extreme case than the W bubble of \xs.  Both nuclear bubbles of NGC~4438 are associated with enhanced radio emission with a projected width of $\sim 3^{\prime\prime}$, which is not well resolved by the existing radio data \citep{Li2022}. 

More complex radio/X-ray bipolar structures have been observed in and around nearby galaxies. A particularly well-known case is the radio lobes and X-ray cones associated with the Seyfert-starburst composite galaxy NGC 3079 \citep[e.g.,][]{veilleux1994,Pietsch1998,Cecil2002,Irwin2012,irwin2019,Li2019,Sebastian2019,Hodges-Kluck2020}. Emanating from the galactic core region of the galaxy, the radio lobes extend $\sim 2$~kpc at 1.4 GHz \citep[e.g.,][]{Sebastian2019} and are not closely correlated with diffuse X-ray emission or other multi-wavelength features. The radio lobes represent only parts of a global outflow dominated by nuclear starburst feedback. In fact, there are also much more extended spur-like or loop-like radio structures mixed in with the large-scale radio halo of the galaxy, especially visible at lower frequencies \citep[e.g.,][]{Irwin2003}. Such extended radio halos with prominent protrusions are not uncommon around starburst galaxies \citep[e.g., NGC 253; ][]{Carilli1992,Heesen2011}. The complex interplay of different energetic processes and outflows makes it difficult to distinguish their respective effects on the observed structures \citep[e.g.,][]{Sebastian2019,Clavijo-Bohorquez2023}. In addition, nuclear radio lobes on sub-kpc scales are also observed in NGC 2992, where the diffuse X-ray emission is detected only in the immediate vicinity of the AGN \citep{Irwin2017,Xu2022,VittoriaZanchettin2023}. The radio and X-ray correspondence is hardly clear. Thus, we focus here on large-scale extraplanar structures that are relatively well-defined in both radio and X-ray and appear to be dominated by AGN feedback.

We are not aware of any other well-defined diffuse radio/X-ray bubbles in nearby disk galaxies that do not appear to be driven primarily by galactic nuclear starbursts. Nevertheless, the above comparison of the three bipolar structures (\sou, \xs\ and NGC~4438) suggests a trend of decreasing plasma temperature with increasing size. The following discussion is aimed at gaining insight into this and possible other trends from comparison with hydrodynamical simulations of such structures. 

\subsection{Comparison with simulations}
\label{ss:dis_comp_sim}
We focus this comparison on galactic bubble structures simulated in a cosmological context. Based on the study of 127 TNG50 simulated Milky Way/M31-like galaxies viewed edge-on, \citet{Pillepich2021} have identified large-scale high-pressure features that are morphologically similar to the eROSITA/Fermi bubbles. Such features are present in about 2/3 of the simulated galaxies and often appear in more or less symmetrical pairs above and below the galactic disks.  Some of the galaxies contain multiple bubbles or shells with increasing sizes away from the galactic disks, resulting from multiple energetic energy releases from the accreting SMBH every 20-50~Myr.  The {\sl global} morphology of the features is not sensitive to the exact form of the energy release (jets and/or winds from hot AGN accretion). However, only a small fraction of the features have extents smaller than 10 kpc. There are hardly any bubbles with extents $\lesssim 5$~kpc, but $\gtrsim $ 2.5~kpc -- the smallest extent at which the identification is made.  
This trend can be understood because the growth of a bubble generally slows down as its size increases -- resulting in an increasing probability of being seen in the simulations.  Therefore, small features like those seen in \xs\ and NGC~4438 are likely to be rare, which could explain why so few features similar to the \xs\ structure have been found in or around other nearby disk galaxies. 

A consequence of the above reasoning is that, if TNG50 is describing a realistic population of bubbles, we should have found numerous "large" (>10 kpc) ones in X-ray observations of nearby galaxies. A possible explanation is, however, that such features are too dim to be detected as the surface brightness of a bubble generally decreases with its extent. While the predicted brightness is typically $\lesssim 10^{35-36} {\rm~erg~s^{-1}~kpc^{-2}}$ for the simulated bubbles (with their extents typically $\gtrsim 10$~kpc), compared to $\sim 4 \times 10^{36} {\rm~erg~s^{-1}~kpc^{-2}}$ for the eROSITA bubbles (extent $\sim 14$~kpc), $\sim 4 \times 10^{38} {\rm~erg~s^{-1}~kpc^{-2}}$ for the \xs\ bubbles ($\sim 8$~kpc), or $\sim 3 \times 10^{40} {\rm~erg~s^{-1}~kpc^{-2}}$ for the northwest nuclear bubble ($\sim 0.3$~kpc) in NGC~4438 \citep{Machacek2004,Li2022}.  Thus it is empirically evident that the surface brightness of a bubble decreases rapidly with its expansion. Another possibility is that bubbles in TNG50 are intrinsically larger and more frequent than those in real galaxies, but this will have to be assessed with better statistics.

In the work of \citet{Pillepich2021}, the plasma temperature is characterized by mass-weighted values, typically in the range of T$\approx10^{6.4-7.2}$~K. This range is not inconsistent with the values given above for the observed features (\S~\ref{ss:dis_comp_obs}). However, it should be noted that the temperatures estimated from the spectral fits are biased toward lower values if the X-ray-emitting plasma is not isothermal. 
Furthermore, our X-ray spectral results show that the mean plasma temperature tends to be lower at the edges than inside the bubbles, especially when its anticorrelation with $\sigma_x$ is taken into account  (Table~\ref{t:t4}). This trend indicates that the average velocity of the shock, responsible for the heating of the plasma, has decreased with time, as expected for a superbubble expanding into the CGM, similar to the scenario proposed for \sou\ \citep[e.g.,][]{Mou2014}. In this case, the ejected material from galactic central regions, most likely solar or supersolar, has been largely diluted by the heated CGM, which may well be sub-solar \citep[e.g.,][]{Ursino2016,Miller2016,Kataoka2018}. 
The metallicity of the simulated bubble is in the range of 0.5-2 Z$_{\odot}$, consistent with $\sim 0.42 Z_{\odot}$ and 0.76 Z$_{\odot}$ for the E and W bubbles within the uncertainties of the estimates (Table~\ref{t:t4}); the emission measured-weighted metallicity is also expected to be biased toward lower values. Lower metal abundances ($\sim 0.2 Z_{\odot}$) are obtained for \sou\ \citep[e.g.,][]{Kataoka2018}. 

In the TNG50 simulations, such energetic bubbles are produced by feedback from AGNs with Eddington ratios of typically $\sim 10^{-5} - 10^{-4}$ \citep{Pillepich2021}. The misalignment between the jets and the general orientation of the bubbles in \xs\ also does not seem to be a problem in the AGN feedback scenario. For example, simulations of \citet{Sarkar2023} show that jets in a disk galaxy tend to be locally strangulated \citep[e.g., via the global 
 kink instability and/or the interaction with highly inhomogeneous ambient medium;][]{Tchekhovskoy2016}, producing high-pressure expanding cocoons. The overall morphology of the resulting large-scale bubbles is mostly determined by the distribution of the surrounding medium (i.e., the ISM and the CGM), not by the direction of the jets.

Therefore, the observations are in broad agreement with the simulations with respect to the X-ray morphology and surface intensity, and the temperature and metallicity of the superbubble structures.  More work will be needed to compare the frequency and size distribution of the real and simulated bubbles.

\subsection{Origin of the \xs\ structure}
\label{ss:dis_origin}
We here specifically discuss what could have driven the \xs\ bipolar structure. To do so, we first estimate the expansion rate ($\dot{R}$) of the radius ($R$), total energy ($E$), and age ($t$) of the structure using the superbubble model \citep[][see also \citet{Miller2016}]{MacLow1988}. Assuming a constant mechanical energy input rate and a uniform surrounding medium, this model gives the following equations:
\begin{equation}
    R=(0.27 {\rm~kpc}) \dot{E}_{38}^{1/5}n_0^{-1/5} t_7^{3/5},
    \label{e:r}
\end{equation}
and 
\begin{equation}
    \dot{R}=(16 {\rm~km~s^{-1}}) \dot{E}_{38}^{1/5}n_0^{-1/5} t_7^{-2/5},
    \label{e:v}
\end{equation}
where $\dot{E}_{38}$ (in units of $10^{38} {\rm~erg~s^{-1}}$), $n_0$ (${\rm cm^{-3}}$), and $t_7$ ($10^7$~yr) are the mechanical energy input rate, external medium number density, and age of a superbubble.

The above two equations can be used to get 
\begin{equation}
    t=(8\times 10^{6} {\rm~yr}) \Big(\frac{R}{4{\rm~kpc}}\Big) \Big(\frac{\dot{R}}{300 {\rm~km~s^{-1}}}\Big)^{-1},
    \label{e:t}
\end{equation}
and
\begin{equation}
    E=\dot{E} t= (3.5 \times 10^{56} {\rm~erg}) \Big(\frac{n_0}{10^{-2}{\rm~cm^{-3}}}\Big) \Big(\frac{R}{4{\rm~kpc}}\Big)^3 \Big(\frac{\dot{R}}{300 {\rm~km~s^{-1}}}\Big)^2,
    \label{e:e}
\end{equation}
where we have assumed a characteristic radius as half the bubble size of the \xs\ bipolar structure.  For the typical expansion velocity we have used 300 km/s estimated as follows.  The thermal energy ($E_{th}$) of the bubble interiors derived from our X-ray spectral fits (Table~\ref{t:t6}) should represent a good fraction of the total energy: i.e., $E_{th} \approx \Big(\frac{5}{11}\Big)E$, according to the superbubble model. Considering the $E_{th}$ values listed for the two bubbles in Table~\ref{t:t6}, which are approximately the same, we estimate that each is produced by a total mechanical energy $E \sim 4 \times 10^{56}$~erg. This, together with Eq.~\ref{e:e}, then gives an estimate of the expansion velocity of the bubbles as
\begin{equation}
    \dot{R}=(3.2 \times 10^2 {\rm~km~s^{-1}}) \Big(\frac{n_0}{10^{-2}{\rm~cm^{-3}}}\Big)^{-1/2} \Big(\frac{R}{4{\rm~kpc}}\Big)^{-3/2} \Big(\frac{E}{4 \times 10^{56} {\rm~erg}}\Big)^{1/2},
    \label{e:ev}
\end{equation}
where $n_0 \sim M_{th}/( \mu m_p V_t) \sim n_e$, if $f_h \sim 1$.  
Thus according to Table~\ref{t:t6}, $n_0 \sim 0.02-0.03 {\rm~cm^{-3}}$ might be a reasonable estimate. In reality, of course, the density of the surrounding gas is expected to be stratified in the direction perpendicular to the galactic disk and to decrease with increasing distance from the galactic plane. At present, the bubbles are likely to expand mainly into the CGM, which has a lower density compared to the bubbles. With this in mind, we assume a characteristic value of $n_0 \sim 0.01 {\rm~cm^{-3}}$ in our parameter estimates. 

Using the above $\dot{R}$ and Eq.~\ref{e:t}, we derive the age of the \xs\ structure to be $t \sim 8 \times 10^6$~yr. This age is smaller than the age ($\sim 3 \times 10^7$~yr) estimated for \sou, assuming a similar formation scenario \citep[e.g.,][]{Mou2014,Miller2016,Sarkar2023}. This is expected because the \xs\ structure is about a factor of $\sim 2$ smaller than \sou\ and $t \propto R^{5/3}$ (if both the energy input rate and the properties of the surrounding medium are similar). Indeed, the mechanical energy input rate for the entire \xs\ structure, $\dot{E} \approx 2E/t \sim 4 \times 10^{42} {\rm~erg~s^{-1}}$, is very similar to that estimated for \sou\ \citep[e.g.,][]{Mou2014,Miller2016,Sarkar2023}.

We can now examine what energy source in \xs\ could be responsible for the above energy input. We can rule out that stellar feedback, mainly via supernovae (SN), plays a major role in the formation of the \xs\ structure. \citet{Ogle2014} estimate a total stellar mass of $8 \times 10^9 {\rm~M_\odot}$ and a star formation rate of $\sim 0.08 {\rm~M_\odot~yr^{-1}}$ in the central 3.4 kpc$^2$ of the galaxy, which is comparable to the corresponding values (within their uncertainties) found in the central molecular zone ( with however a much smaller size of $\sim 300$~pc) of our Galaxy \citep[e.g., ][]{Henshaw2022}. Therefore, the stellar feedback mechanical energy input rate from the central region of \xs\ should be similar to that ($(0.6-5)\times 10^{40} {\rm~erg~s^{-1}}$) estimated from the Galactic central molecular zone, assuming $10^{51}$~erg per SN \citep{Crocker2011b,Ponti2015}. This rate is grossly insufficient to produce the \xs\ structure if our estimate of its age is realistic. The stellar feedback scenario also has difficulty explaining the discrete nature of the bubbles. \xs\ is clearly not a nuclear starburst galaxy; in fact, the study by \citet{Ogle2014} suggests that much of the gas in the central region of the galaxy has been ejected into the halo by recent AGN.  Nor have similar energetic radio/X-ray structures been observed in extensive surveys of nearby highly inclined disk galaxies with overall comparable or slightly higher star formation rates \citep[e.g.][]{Wiegert2015,Li2013,Li2017}. 

We next consider AGN  feedback as the energy source of the \xs\ structure. The presence of a low luminosity AGN in the galaxy is well established \citep[e.g.,][]{Makishima1994,Lasota1996,Yuan2002,VronCetty2006,Masini2022}. Its SMBH is surrounded by a molecular disk in Keplerian motion, as traced by masers observed with very long baseline interferometry. The measured mass of the black hole is $4 \times 10^7 {\rm~M_\odot}$ \citep{Nakai1993,Miyoshi1995}. The recent study by \citet{Masini2022} shows that the X-ray properties of the AGN fluctuate on timescales of hours to years, both intrinsically and due to absorption. 
The 2-10 keV intrinsic luminosity was $\sim 10^{41} {\rm~erg~s^{-1}}$ in the early 2000s, but decreased by a factor of $\sim 3$ by 2016. The estimated bolometric correction is $\sim 20$. 
The low Eddington ratio ($\sim 10^{-4}$) of the AGN is well within the regime expected for a radiatively inefficient accretion flow, which could release much of the mechanical energy via jets \citep{Yuan2002}. The current power of the jets is estimated to be in the range of a few times $(10^{39}-10^{41}) {\rm~erg~s^{-1}}$ \citep[e.g.,][]{Krause2007}. The jet power is needed to heat the surrounding gas of different phases, to drive the gas motion, and to accelerate cosmic rays. While the present jets alone are not powerful enough to produce the observed bipolar structure in \xs. 

One possibility is that the wind from the AGN plays a major role in producing the \xs\ structure. Strong outflows or winds are the ubiquitous phenomena of AGNs, as shown in observations and in numerical simulations \citep[e.g.,][]{Yuan2012,Yuan2015,King2015}. Driven by the combination of the centrifugal force and the magnetic pressure gradient associated with the accretion flows, such winds have much larger opening angles compared to jets and are generally more effective in coupling with the surrounding ISM \citep{Ostriker2010,King2015}. This low luminosity AGN scenario for the \xs\ structure is quite similar to that advocated for \sou\ in recent years \citep[e.g.,][]{Mou2014,Sebastian2019,Sarkar2023}.

The AGN could also have been much more powerful in the recent past. We consider an extreme case in which the structure was initially produced by an instantaneous energy input and has undergone an adiabatic expansion. In this case, the expansion of a bubble can be described by \citep[e.g.][]{Woltjer1972}
\begin{equation}
R \sim 1.17  \Big(\frac{E}{n_0\mu m_p}\Big)^{1/5}t^{2/5},
\label{e:r2}
\end{equation}
{where E is the energy resealed, n$_0$ is the number density of the external medium, $\mu$ is the atomic number and m$_p$ is the proton mass.}

Assuming that $E$ is  currently dominated by the thermal energy $E_{th} \sim 2 \times 10^{56}$~erg for each of the two bubbles of the structure (Table~\ref{t:t6}) and again $n_0 \sim 10^{-2} {\rm~cm^{-3}}$,
we estimate $\dot{R} \sim 430 {\rm~km~s^{-1}}$ and $t \sim 3.7 \times 10^6$~yr.
Reality should lie between these two extremes (constant or instantaneous energy input). We, therefore, expect the current expansion velocity and age of the structure to be $\sim (320-430) {\rm~km~s^{-1}}$ and $\sim (4-8) \times 10^6 $~yr, respectively.

We may use the above expansion velocity to estimate the shock temperature to be $kT \sim \frac{3}{16} \mu m_p \dot{R}^2=0.18$~keV (assuming a strong shock). This value is comparable to or slightly smaller than the temperatures from our spectral fits for the bubble edges (Tables~\ref{t:t4}), suggesting that the shock may actually be quite weak (probably moving mostly in a hot CGM of comparable temperature) and explaining the lack of a sharp boundary around the \xs\ structure on the northeast and southwest sides. 
The generally hotter plasma inside the bubbles is expected because its heating occurred earlier when the bubble expansion velocity was greater. 
These apparent consistencies lead us to strongly favor the
recent AGN origin of the \xs\ structure.

Although the global morphology of the structure tends to be largely determined by the density gradient of the ISM/CGM, the distribution of the observed radio or X-ray emission may still be strongly influenced by the specific dissipation process of the jets. The presence of the anomalous arms in the galaxy is attributed to the large misalignment of the jets with the rotational axis of \xs, as suggested by the present offset \citep[$\sim 120^\circ$; ][]{Krause2007} between the spin vectors of the nuclear disk and the galactic disk. It has also been proposed that the jets have precessed through the galactic gaseous disk, causing their strong dissipation or energy deposition in the ISM, probably via a series of oblique shocks, which may be responsible for the S-shaped intensity enhancement of the bipolar structure \citep{Cecil2000}. Other relative motions of the jets with respect to the surrounding medium (e.g. due to the rotation of the galactic disk) may also be important. In any case, the interaction of a jet with respect to the medium is naturally expected to create a high-pressure cocoon that is only partially confined by the ram pressure due to the relative motions. 
Since the medium may be highly inhomogeneous, the degree of both the interaction and the confinement is expected to vary greatly from time to time, resulting in different penetrations of the jet through the disk and heating of the gas to different temperatures. Particles from the jets may also be (re)accelerated. The high-pressure materials from the dissipated jets and the heated ambient medium (both thermal and non-thermal) inevitably drive flows and potential bifurcations into multiple streams if dense obstacles are encountered. 

\section{Summary}\label{s:sum}
We have presented the discovery and analysis of a pair of radio/X-ray bubbles located above and below the disk of the nearby disk galaxy M106, using the recently released \rins\ survey data in the 120-168 MHz and 42-66 MHz bands and the Chandra data archive. Our study includes spatial and spectral analyses of the radio and X-ray data to understand the properties of the bubbles and compare them with the observations of \sou\ and to cosmological simulations. Our main findings are summarized as follows:
\begin{itemize}
\item The bipolar structure of \xs\ includes the diffuse radio and X-ray emissions from the bubbles and their two unilaterally enhanced edges (Fig.~\ref{fig:f2}). These edges were previously known as two anomalous arms of the galaxy, and together they form an S-shaped structure, which is quite different from the one-sided (Galactic northeast) enhancement (i.e., the Radio Loop I and X-ray NPS) observed in \sou. 
The \xs\ structure extends $\sim 8$~kpc away from the galactic disk of M106 and is a factor of $\sim 2$ smaller in size than the eROSITA bubbles. 
\item The radio and X-ray emissions from the \xs\ structure show a general morphological similarity, suggesting that they are physically associated or coexist in the bubbles. However, there is a lack of detailed correspondence between radio and X-ray substructures, indicating that they can be produced independently. Some of the substructures (e.g., the E hot spot) may just represent features overlapping in projection with the bubbles.
\item The radio spectral index maps obtained using the LOFAR bands 
clearly show that the radio emission from the structure is non-thermal and thus synchrotron in origin. The index has a mean value of $\sim 1$ over the bulk of the bubbles. Assuming the energy equipartition between cosmic rays and magnetic fields, we estimate the mean magnetic field strength to be $\sim 3-4\,\mu$G at the bubble centers and up to $\sim 15\,\mu$G at the edges. However, the field strength can be more than an order of magnitude higher in and around radio emission peaks within the edges that cannot be resolved by the LOFAR data. 
\item The spectral index and the intensity of the radio emission further show an anti-correlation in the edges, as seen in higher-resolution radio images. This anti-correlation is naturally explained by the synchrotron cooling of the cosmic-ray electrons on time scales of $\sim 1 \times 10^6$~yr at $\nu_c \sim 54$~MHz.
\item The X-ray spectra of the bubbles can be reasonably well characterized as an optically thin thermal plasma with a lognormal temperature distribution with a mean at $\sim 0.2$~keV and a dispersion $\sigma$(lnT) $\sim 1$. The plasma temperature tends to decrease with the increasing distance from the galaxy's disc of the galaxy. The E/W bubble interiors have luminosities of $\sim 4.7/7.7 \times 10^{39}$ erg s$^{-1}$. The total thermal energy is $\sim 3.3 \times 10^{56}$ erg, while the mean cooling time scale of $\sim 1$~Gyr. The thermal plasma seems to dominate the pressure and thus the force driving the expansion of the bubbles. The metal abundance of the plasma appears to be sub-solar, suggesting that much of the plasma is the heated CGM of relatively lower metallicity, diluting the chemical enrichment of the ejected material from the galaxy's central region. 
\item Our results are broadly consistent with those expected from the TNG50 simulations \citep[e.g.,][]{Pillepich2021} 
 and indicate that the \xs\ structure is the result of AGN feedback. Indeed, we find that the energetics of the stellar feedback in the central region of \xs\ is far from sufficient to produce the structure, which has a characteristic age of $\sim 8 \times 10^6$~yr and requires a mechanical energy input at the average rate of $\sim 4 \times 10^{42} {\rm~erg~s^{-1}}$. However, the current jets of the AGN do not seem powerful enough to give this input. Most likely, the AGN was substantially more energetic in the recent past and/or has released a wind that is significantly more powerful than the jets.
 \item While a large-scale bipolar structure can be inflated by the collective energy deposition of an AGN in the ISM/CGM with the expected density distribution, morphological asymmetry as observed in the radio/X-ray emissions of \xs\ can be naturally explained by a strong misalignment of its AGN jets from the rotational axis of the disk galaxy. This explanation may also apply to asymmetric morphology observed in other similar structures, including \sou.
\item Older and larger structures may be present in the galaxy's CGM. The Chandra data already show evidence of diffuse X-ray emission from the galaxy on scales larger than the bipolar superbubble structure. However, the counting statistics and the spatial coverage of the data are too limited to allow a detailed analysis of this large-scale hot CGM and its substructures.
\item The \xs\ structure is apparently a younger version of \sou. These two structures are probably similarly powered and have evolved in a comparable galactic environment. But the \xs\ structure allows a view with considerably less or different confusion with the galactic disk material.
\end{itemize}

 While the present work was initially motivated by the striking appearance of the radio bubbles of \xs\ in the \rins\ DR2 release, more about the presence of galactic nuclear outflow structures could be revealed by a systematic survey of the X-ray data archive (including the release of the eROSITA sky survey data in the near future), as well as the \rins\ data in both the 120-168 MHz and 42-66 MHz bands. Future deeper high-resolution observations with broader field coverage may further reveal larger-scale dome-like or cocoon-like features around nearby disk galaxies, as seen in some cosmological hydrodynamical simulations \citep[e.g.,][]{Pillepich2021}, providing important insights into the role of galactic feedback in regulating the structure and evolution of the CGM.

 \section*{Acknowledgment}
We thank the referee, as well as Raffaella
Morganti, for their constructive comments, which helped to improve the paper. We acknowledge the help of Luan Luan and Yang Yang who provided guidance for a portion of the data analysis. This work is partly based on data obtained from the Chandra Data Archive and software provided by the Chandra X-ray Center in the application packages CIAO and Sherpa and on \rins\ data products provided by the \rins\ Surveys Key Science project (LSKSP; https://lofar-surveys.org/) and derived from observations with the International \rins\ Telescope, 
which is collectively operated by the ILT foundation under a joint scientific policy. The efforts of the LSKSP have benefited from funding from the European Research Council, NOVA, NWO, CNRS-INSU, the SURF Co-operative, the UK Science and Technology Funding Council, and the J\"{u}lich Supercomputing Centre.

 \section*{DATA AVAILABILITY}

The X-ray data on M106 as described in Section 2 include the two ACIS-S observations listed in Table 2 and are available in the Chandra data archive (https://asc.harvard.edu/cda/), while the \rins\ data 
are available at https://lofar-surveys.org/surveys.html.
Processed data products underlying this article will be shared on
reasonable request to the authors.




\bibliographystyle{mnras}
\bibliography{example} 
\bsp	

\appendix 
\section{Spectral analysis of the local X-ray background}\label{a:spec_bkg}
We fit the local background spectrum with a model comprising various expected components (Eq. \ref{Eq:back}):
\begin{equation}
  \mathrm{APEC_{LB}+TBABS(APEC+POWERLAW)}, \label{Eq:back}
\end{equation}
where APEC is the plasma model same as the VAPEC, but with the metal abundances fixed to the default solar values in Xspec.
The APEC model represents collisional-ionized plasma and depends on four parameters: temperature, abundance, redshift, and normalization. Here we fit the parameters for temperature and normalization only. The TBABS model characterizes the foreground hydrogen column density. The POWERLAW model is a standard power law parameterized by the photon index and normalization.
The APEC$_\mathrm{LB}$ component in our model represents the contribution from the Local Bubble with a fixed temperature of 0.1 keV \citep{Smith2001}., while the combination of APEC and POWERLAW characterizes the background emission from more distant contributions from diffuse hot gas and unresolved point-like sources, chiefly AGNs. This emission is subject to the Galactic foreground absorption (TBABS) with a hydrogen column density of N$_{H}=4.2\times 10^{20}$ cm$^{-2}$ in the direction of \xs\ \citep{HI4PICollaboration2016}. The fitting is satisfactory ($\chi^2/dof=20/22$) and gives the Xspec normalizations of the as 0.246$_{-0.03}^{+0.05}$, 1.73$_{-5.57}^{+7.87}$, and 8.82$_{-3.00}^{+2.96}$ for the APEC$_\mathrm{LB}$, APEC and power law, as well as the APEC temperature as 4.54$_{-3.37}^{+2.65}$.
\label{lastpage}
\end{document}